\documentclass[pre,aps,10pt,twocolumn,floatfix,superscriptaddress]{revtex4-1}
\usepackage[utf8]{inputenc}
\usepackage[english]{babel}
\usepackage{graphicx}
\usepackage{amsmath}
\usepackage{amssymb}
\usepackage{dcolumn}

\begin{document}

\title{Quantum shell effects in compressed mesoscopic system}
\author{S.E. Kuratov}
\affiliation{Dukhov Research Institute of Automatics (VNIIA), Sushchevskaya str. 22, Moscow, 127055, Russia}
\author{D.S. Shidlovski}
\author{S.I. Blinnikov}
\affiliation{Dukhov Research Institute of Automatics (VNIIA), Sushchevskaya str. 22, Moscow, 127055, Russia}
\affiliation{Institute for Theoretical and Experimental Physics, Bol’shaya Cheremushkinskaya str. 25, Moscow, 117218, Russia}
\email{ser.evg.kuratov@gmail.com, dmitry.shidlovski@yandex.ru}

\begin{abstract}
The article demonstrates the nontrivial manifestation of quantum shell effects in a compressed mesoscopic system.
It is shown that there are two spatial scales in the distribution of degenerate electrons in a spherical well.
The first scale is the Fermi length $\sim h/p_{\rm F}$. By quantum shell effect, the authors mean the
existence of the new spatial scale, which is order of the system size and much larger than the first
scale.
The theoretical analysis for the large amount of free electrons ($N \lesssim 10^9$) in an infinite spherical
well demonstrates what causes the appearance of the spatial nonuniformity and gives analytical
expression for the electron distribution function.
These results are confirmed by a numerical summation of exact solutions for the electron wave functions in an infinite potential well.
It is shown that an analogous effect for the spatial distribution of electrons exists in a compressed
hydrogen gas bubble of submicron size ($<0.1 \mu m$).
The numerical simulation of the electron distribution was carried out by the DFT (Density Functional
Theory) method.
The consequence of this effect is the nontrivial dynamics of the compressible cold gas bubble.
This system can be realized in the thermonuclear experiments. The limiting factors of the analyzed effect are considered: symmetry of system, electron temperature, and curvature of system boundary.
\\

Keywords: quantum shell effects, compressed mesoscopic system, spatial distribution, hydrogen gas bubble, DFT method.\\

{\noindent PACS number(s): 03.75.-b, 71.15.Mb, 36.40.-c, 71.10.Ca}
\end{abstract}

\maketitle

\section{Introduction}

There are numerous manifestations of orbital quantum effects in different systems: shell model of the nucleus, orbital effects in the atomic spectrum, oscillatory behavior of the energy spectrum of nanoclusters.

It is known, that the density of the electron distribution has the oscillating behavior because of the presence of shell corrections in a spherically symmetric potential~\cite{Kirzhnits1975}. The spatial scale is the order of the atomic size. In a spherically symmetric potential the density of the electron distribution often shows an oscillating behavior along the radius.
This can be seen in the solutions by Hartree-Fock, Hartree-Fock-Dirac methods, in Thomas-Fermi approximation~\cite{Kirzhnits1975,Shpatakovskaya2012}, and in DFT~\cite{Ekardt}.
Such oscillating behavior is observed, not only for the electrons, but also for the other particles,
and for Coulomb as well as for other interaction potentials between particles (Yukawa potential, hard core, etc.)~\cite{PhysRevE.85.056402}. In the atom, the number of electrons is relatively small ($N<10^2$).

In the present article we show that the nonuniformity of the distribution is also present for a macroscopic number of electrons ($N \lesssim 10^9$) in a potential well.

The theoretical analysis of the free electron distribution in a spherical potential well
shows what causes the appearance of the nonuniformity in the system.
It also demonstrates that the spatial scale of the nonuniformity is of the same order of the well radius and can be greater than the distance between the particles by several orders.

We present also the effective computational technique that allows one to calculate the electron density by numerical summation of exact solutions for the wave functions of free electrons in a potential well.
We perform the calculations of various systems with the electron number of up to 1 billion particles.
These results confirm the existence of the effect and analytical expression obtained for the electron distribution function.

We analyze the inhomogeneous spatial distribution of electrons in the strongly compressed gas bubble (H$_2$, D$_2$) of submicron size. The characteristic values of the thermodynamic quantities of the compressed gas are the following: $\rho_{\rm gas} \sim$ (10-30) g/cm$^3$, $N_e \sim$ 10$^{30-31}$ m$^{-3}$, $E_F \sim$ (20-100) eV, $T_i \sim T_e \sim$ (0.1-1) eV. All electrons are ionized and degenerate, and the ions are the classical nonideal gas.

The numerical simulation of the electron distribution was carried out by the DFT (Density Functional
Theory) method.
We used the jellium model for hydrogen clusters (the number of atoms varies from 4000 to 100,000).
Obtained results confirm the existence of the analogous effect in the compressed hydrogen gas bubble.

Since the scale of the inhomogeneity of the electron distribution is much larger than the interatomic
distance, the effect is manifested in hydrodynamic relaxation processes of the ion system and can be
observed in experiments. We analyze two problems: the hydrostatic equilibrium of a
compressed gas bubble and the compression dynamics of the gas bubble.
The analysis indicates a nontrivial
dynamics of gas compression, which is fundamentally different from the process of adiabatic compression
in traditional systems.

In the final part of the article, we analyze the factors limiting the manifestation of the effect.
The main ones are the symmetry of the system, the electronic temperature and the curvature of the boundary of the system.

\section{System of free degenerate electrons in an infinite spherical well}

The existence of the nonuniformity of the distribution can be demonstrated by the example of the simplest problem: a one-dimensional electron gas in an infinite well. The wave functions of electrons have the form
\begin{equation}
 \Psi_k=\sqrt{\dfrac{2}{L}}\sin\dfrac{\pi kx}{L},\quad E_k=\frac{\pi^2\hbar^2k^2}{2m_eL^2}.
\end{equation}
And the concentration $n$ is determined by the expression
\begin{multline}
 n=\sum_1^{N_0}\Psi_k^2=
 \sum_{k=1}^{N_0}\left(\sqrt{\dfrac{2}{L}}\sin\dfrac{\pi kx}{L}\right)^2=\\
 =\dfrac{1}{L}\sum_{k=1}^{N_0}\left(1-\cos\dfrac{2\pi kx}{L}\right)=\\
 =n_0\left(1-\dfrac{1}{N_0}\csc\dfrac{\pi x}{L}\sin\dfrac{(N_0+1)\pi x}{L}\cos\dfrac{N_0\pi x}{L}\right).
\end{multline}
The final expression for concentration $n$ has two spatial scales. The first scale is the distance between the particles $L/N_0$ [the factors $\sin\left((N_0+1)\pi x/L\right)\cos\left(N_0\pi x/L\right)$], the second scale is the order of the system size $L$ [the factor $\csc(\pi x/L)$]. In the considered flat one-dimensional system, the effect is negligible
\begin{equation}
\label{delta_n_flat}
 \left(\frac{\Delta n}{n_0}\right)_{\rm flat} \sim \frac{1}{N_0}.
\end{equation}

In the present paper, this effect is studied for the electrons in a spherical potential well.
We show that, due to the quantum shell effects, the magnitude of the inhomogeneity in the
electron distribution increases significantly
\begin{equation}
\label{delta_n_sph1}
 \left(\frac{\Delta n}{n_0}\right)_{\rm Sph} \sim \frac{1}{\sqrt{N_0}}.
\end{equation}

For the large amount of electrons ($N \sim 10^9$), these quantities~\eqref{delta_n_flat} and~\eqref{delta_n_sph1} can differ by several orders of magnitude.

\subsection{The theoretical analysis}
At first we analyze the distribution of free semiclassical electrons in an infinite spherical well. All electrons are degenerate in this system and their Fermi energy is greater than the thermal and Coulomb energies.
Moreover, the electrons have the significant orbital angular momentum and thus they have semiclassical behavior.
Therefore such a formulation of the problem is close to the real situation.
The solution of this problem gives us the analytical dependences of the electron distribution on the system parameters.

The semiclassical approach was widely used for the analysis of metal clusters~\cite{Shpatakovskaya2012}, for the calculation of the nuclei energy spectrum~\cite{Strutinsky}, for the calculation of the electron concentration oscillations in the atom.

To determine the electron concentration, we employ the Green's function representation $G (r'',r',e)$ for the electrons in the semiclassical approximation~\cite{Strutinsky}.

\begin{widetext}
\begin{equation}
\label{green_function}
G \left ( r'', r', e \right ) = G_0 - \frac{1}{(2 \pi h^5)^{1/2}}
\sum_{\alpha}\left\{ p_{\rho} D^{1/2} \exp \left( \displaystyle \frac{i}{h} S_{\alpha} \left ( r'',
r', e \right) + i \nu - \frac{i \pi}{4} \right) \right\}_{\alpha}
\end{equation}
\end{widetext}

$S_{\alpha}$ - classical action integral
\begin{equation}
S_{\alpha} = \int_{r'}^{r''} p_{\alpha} dl_{\alpha}
\end{equation}
\begin{equation}
G_0 \left ( r'', r', e \right ) = - \frac{m}{2 \pi h^2 | r'' - r' |}
\exp \left(\displaystyle \frac{i}{h} | r'' - r' | p(r) \right)
\end{equation}
\begin{equation}
r = \frac{r' + r''}{2}, p' = \frac{dS_{\alpha} (r', r'', e)}{dr'}, p'' = \frac{dS_{\alpha} (r', r'', e)}{dr''}
\end{equation}

\begin{equation}
D = \det \left [ \begin{array}{ccc}
\displaystyle \frac{dp'_{\rho}}{d\rho''}  & \displaystyle \frac{dp'_{\rho}}{dz''}  &
\displaystyle \frac{dp'_{\rho}}{de''} \\
&& \\
\displaystyle \frac{dp'_{z}}{d\rho''}     & \displaystyle \frac{dp'_{z}}{dz''}     & \displaystyle \frac{dp'_{z}}{de''}    \\
&& \\
\displaystyle \frac{dt_{\alpha}}{d\rho''} & \displaystyle \frac{dt_{\alpha}}{dz''} & \displaystyle 0                       \\
\end{array} \right ]
\end{equation}

\begin{figure*}[t]
  \begin{center}
    \includegraphics[width=0.56\linewidth]{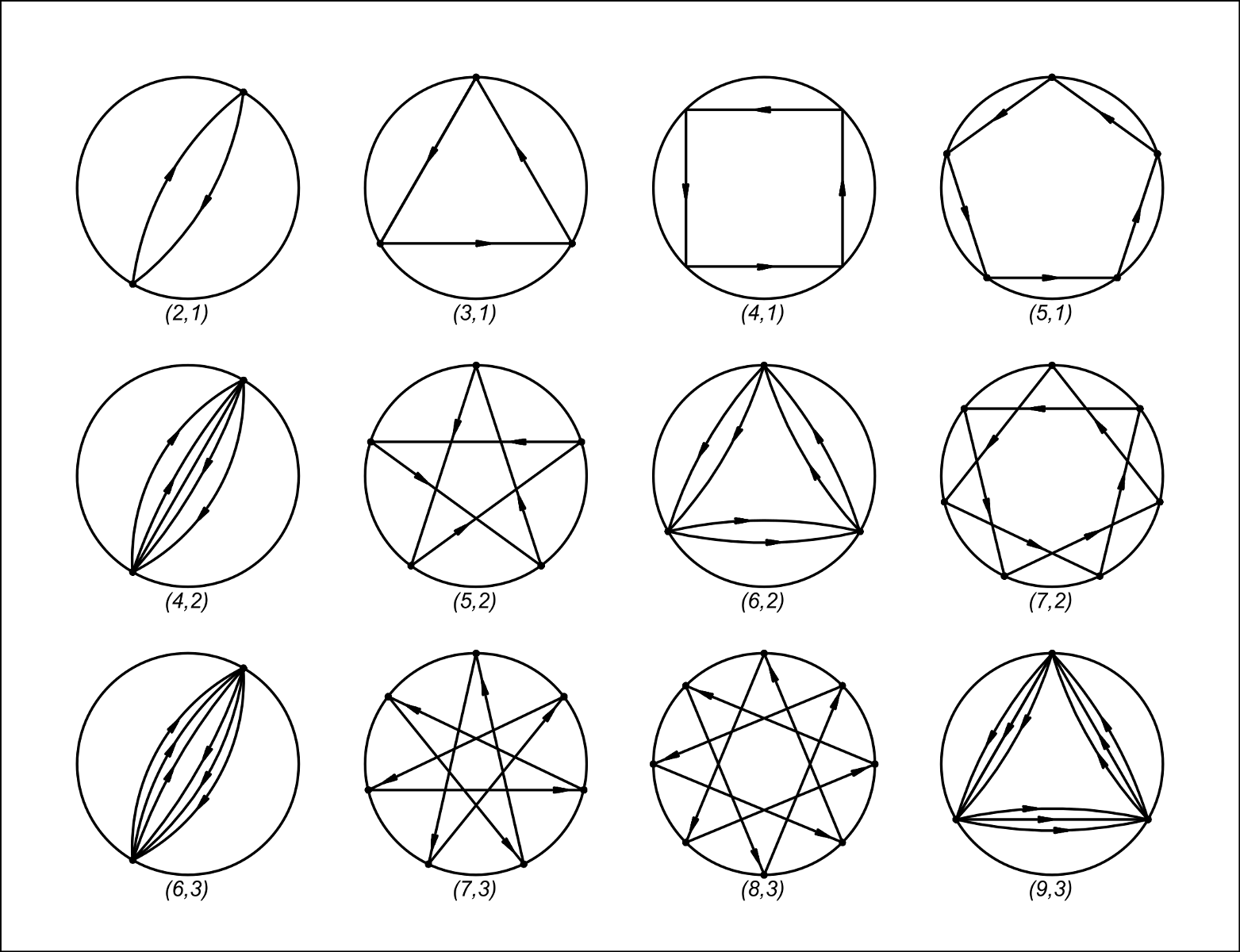}
    \includegraphics[width=0.43\linewidth]{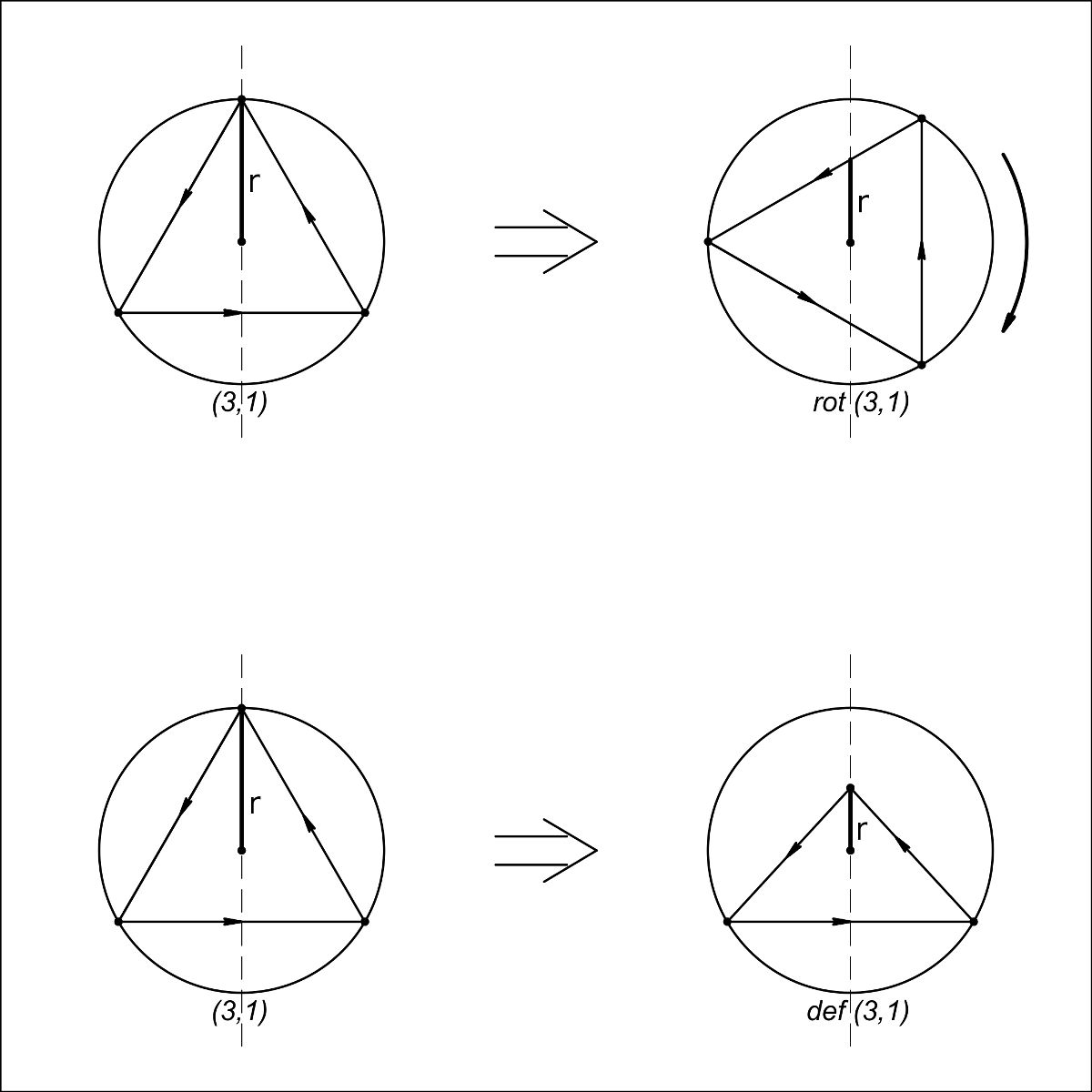}\\
    \hspace{0.08\linewidth}(a)\hspace{0.49\linewidth}(b)\\
    \caption{\label{fig:trajectories} Periodic trajectories of the electron (a). The ways of forming two types of trajectories (b).}
  \end{center}
\end{figure*}

\begin{equation}
t_{\alpha} \left ( r^{\prime}, r'' \right ) = \frac{dS_{\alpha} \left( r',
r'', e \right )}{de} = \int_{r'}^{r''}
\frac{1}{|\dot{\mathbf r}|}
dl_{\alpha}
\end{equation}

In~\eqref{green_function} the summation is extended over all real classical trajectories
$\alpha$ that connect $r'$ and $r''$, $t_{\alpha}$ is the time of the motion along the trajectory $\alpha$.

The electron concentration is determined from the following expression involving imaginary part
of the Green's function
\begin{equation}
\label{density_integral}
n(r) = - \frac{2}{\pi} \int_{-\infty}^{\epsilon_F} \Im \{ G(r, r, e) \} de
\end{equation}

In our case, the calculation of the electron concentration [i.e., integral \eqref{density_integral}],
is reduced to
taking into account a set of orbits which have coincident initial and final coordinates.
There are two types of orbits.
The examples of such trajectories are shown in Fig.~\ref{fig:trajectories}. Such
trajectories are formed from the known periodic orbits (Fig.~\ref{fig:trajectories}~a) in two ways
(Fig.~\ref{fig:trajectories}b).

The first type of trajectory is formed from the periodic orbits by rotation. The method is shown in~Fig.~\ref{fig:trajectories}b by the example of rotation of a triangular periodic orbit $(3.1)$. All trajectories formed by this method will be denoted as $ \mbox{rot} (n, m)$.

The second type of trajectory is formed from the periodic orbits by deformation, see~Fig.~\ref{fig:trajectories}b. Such trajectories will be denoted as $\mbox{def} (n, m)$.

We obtain a general expression for the contribution of each trajectory to the concentration value $n$~\eqref{density_integral}. In our case
\begin{equation}
S_{\alpha} \sim p R_0 L_{\alpha} \left ( \frac{r'}{R_0}, \frac{r''}{R_0} \right ),
\end{equation}

where $L_{\alpha}$ the dimensionless length of the trajectory $\alpha$, $R_0$ -- well radius. We define the dimensionless function $F_{\alpha}$ according to the following expression
\begin{equation}
p_{\rho}D_{\alpha}^{1/2} =\frac{m\sqrt{p}}{\sqrt{R_0}}F_{\alpha} \left(\frac{r'}{R_0},\frac{r''}{R_0}\right).
\end{equation}

The expression for second part of the Green function~\eqref{green_function} has the following form:
\begin{multline}
\label{green_func_expr}
\Delta G(r'', r', e) = \frac{1}{h^{5/2}} p \left ( \frac{m^2}{R_0 p} \right )^{1/2}
F\left ( \frac{r'}{R_0}, \frac{r''}{R_0} \right ) \\
\times\exp \left( \displaystyle \frac{i}{h} p R_0 L_{\alpha} \left ( \frac{r'}{R_0}, \frac{r''}{R_0} \right ) \right),
\end{multline}

\begin{equation}
n(r)-n_0 = \Delta n(r) = -\dfrac{2}{\pi} \int_{-\infty}^{\varepsilon_F} \Im \{\Delta G(r, r, e)\} de.
\end{equation}
For the integral calculation, we use the large value ($\sim10^3$) of the exponent in~\eqref{green_func_expr}. We obtain the following:
\begin{equation}
\begin{split}
\label{delta_n}
\frac{\Delta n(r)}{n_0} = \sum_{\alpha} \frac{F_{\alpha}\left(\dfrac{r}{R_0},\dfrac{r}{R_0} \right )}{L_{\alpha} \left( \dfrac{r}{R_0}, \dfrac{r}{R_0} \right)} \left( \frac{8}{N} \right )^{1/2}\\
\times\sin \left[ \displaystyle \frac{1}{h} p_{\rm F} R_0 L_{\alpha} \left ( \frac{r}{R_0}, \frac{r}{R_0} \right ) \right].
\end{split}
\end{equation}
The contribution of each trajectory in~\eqref{delta_n} is taken additively into account. A number of corollaries follows from the resulting expression.

The dependence of the relative deviation of the concentration on the particle number has the form
\begin{equation}
\label{delta_n_sph}
 \left(\frac{\Delta n}{n_0}\right)_{\rm sphere} \sim \sqrt{\frac{8}{N}},
\end{equation}
Expression~\eqref{delta_n} has two spatial scales. The first scale is the Fermi length $\sim h/p_{\rm F}$, the second scale is the size of the potential well $R_0$.
For the two types of trajectories $\mbox{rot}(n, m)$ and $\mbox{def}(n, m)$ introduced above, the spatial dependence in~\eqref{delta_n} has qualitatively different character.

For the trajectories $\mbox{rot}(n, m)$, the spatial dependence of each term in~\eqref{delta_n} is determined only by the factor $F_{\alpha}$, since $L_{\alpha}$ does not depend on r.
Therefore the total contribution to~\eqref{delta_n} of this type of trajectory forms a function having
only the scale of spatial inhomogeneity of the order of $R_0$. For the trajectories $\mbox{def}(n, m)$, each term of the sum is the product of the rapidly oscillating function $\sin \left[ \displaystyle \frac{1}{h} p_{\rm F} R_0 L_{\alpha} \left( \frac{r}{R_0}, \frac{r}{R_0} \right) \right]$ with the spatial scale of the inhomogeneity $\sim h/p_{\rm F}$ and a slowly varying function $\left. F_{\alpha}\left(\frac{r}{R_0},\frac{r}{R_0} \right )\right/ L_{\alpha} \left( \frac{r}{R_0}, \frac{r}{R_0} \right)$ with a spatial scale $\sim R_0$.
Therefore the relative deviation of the concentration consists of the oscillation and non-oscillatory parts
\begin{equation}
 \left(\frac{\Delta n}{n_0}\right)_{\rm Sph} = \left(\frac{\Delta n}{n_0}\right)_{\rm Oscil} + \left(\frac{\Delta n}{n_0}\right)_{\rm Nonoscil}.
\end{equation}

The first part is determined by the sum~\eqref{delta_n} along the trajectories of the type $\mbox{def}(n, m)$, the second part is determined by the sum~\eqref{delta_n} along the trajectories of the type $\mbox{rot}(n, m)$.

Each term of the sum~\eqref{delta_n} is proportional $1/L_{\alpha}$ and decreases with increasing $m$, since $L_{\alpha}\sim m$.
Therefore the sum is determined by terms with small $m$, i.e., trajectories $\mbox{def}(n, 1)$, $\mbox{rot}(n, 1)$.
The Appendix~\ref{app_trajectories} presents the results of the calculations of $F_{\alpha}$ and
$L_{\alpha}$ for $\mbox{def} (3, 1)$, $\mbox{rot} (3, 1)$, for which it is possible to obtain closed
analytic expressions. The obtained results make it possible to draw general conclusions about the
structure of the expression~\eqref{delta_n} and the functions $F_{\alpha}$ and $L_{\alpha}$.

The non-oscillating part is a part of the sum~\eqref{delta_n}, where $L_{\alpha}$ is constant. The oscillating part is a part of~\eqref{delta_n}, where $L_{\alpha}$ is a function. For $F_{\alpha}$ we have
\begin{equation}
 F_{\alpha}\left(\frac{r}{R_0},\frac{r}{R_0} \right)=\Theta(r-s_{\alpha}) G_{\alpha}\left(\frac{r}{R_0},\frac{r}{d_{\alpha}} \right),
\end{equation}
where $\Theta$ is the Heaviside step function.
The $d_{\alpha}$ is the point of the function $F_{\alpha}$ singularity (see Appendix~\ref{app_trajectories}).

Due to the deviation of the electron concentration from the average value, an electric field appears in the system. We calculate the emerging potential for the oscillating and non-oscillating parts of $\Delta n$ by means of expression
\begin{equation}
\phi(r) = \frac{4\pi e\displaystyle \int_0^r\Delta n(r)r^2 dr}{4\pi\varepsilon_0 r}.
\end{equation}

First, we calculate the contribution of the oscillating part in~\eqref{delta_n}.
We use the stationary phase method. There are two possible cases. In the first case the function $L_{\alpha}$ has no stationary points on the interval of integration, so the potential has the obvious oscillating behavior. Therefore the spatial distribution of the electron has one inhomogeneity scale. This scale is the Fermi length $p_Fr/h$.

\begin{figure}[t]
	\begin{center}
		\includegraphics[width=0.7\linewidth]{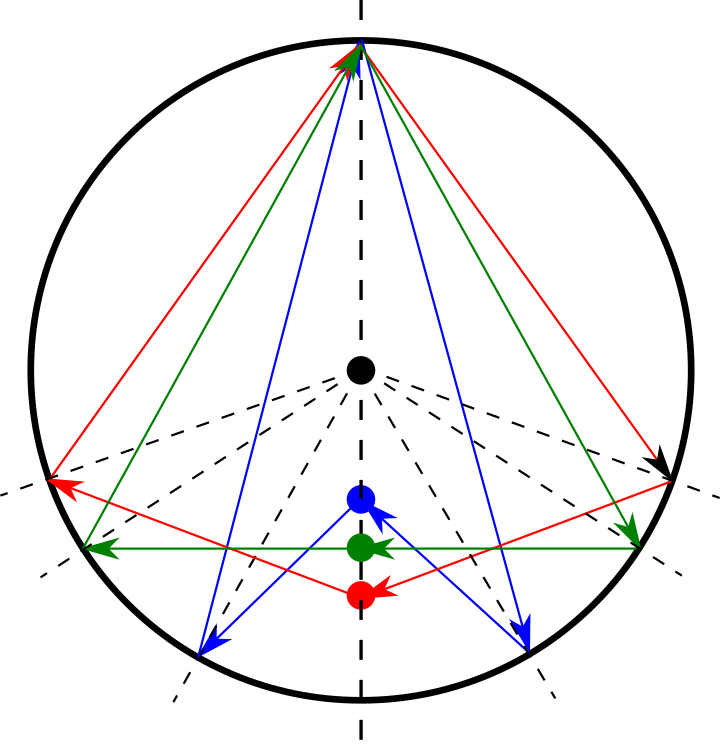}
		\caption{\label{fig:trajectories2} The trajectories of electron $\mbox{def}(4,1)$. The triangle trajectory has a minimal length.}
	\end{center}
\end{figure}

In the second case, the function $L_{\alpha}$ has stationary points. All trajectories of
the type $\mbox{def} (n, m)$ have this feature.
The example of such a trajectory $\mbox{def}(4,1)$ is presented in Fig.~\ref{fig:trajectories2}.

In the latter case, to calculate the potential, we use the known relation of the stationary phase method
\begin{equation}
 \int\limits_a^b f(x)\exp(i\lambda s(x))dx\cong f(x_0)\sqrt{\frac{2\pi}{\lambda s(x_0)''}}\exp(i\lambda s(x_0)),
\end{equation}
$x_0$ is an extremum point, where $s'(x_0)=0$.

We obtain the following expressions for the potentials associated with $\Delta n_{\rm oscil}$ and $\Delta n_{\rm nonoscil}$:
\begin{widetext}
\begin{equation}
\label{delta_v_osc}
 \Delta \varphi_{\rm oscil} = \left(\frac{18}{\pi^2} \right)^{1/3}
 \frac{eN^{1/3}}{\varepsilon_0r} \sum_{\alpha}
 \frac{F_{\alpha}\left(\dfrac{d_{\alpha}}{R_0},\dfrac{d_{\alpha}}{R_0} \right)}
 {L_{\alpha} \left( \dfrac{d_{\alpha}}{R_0}, \dfrac{d_{\alpha}}{R_0} \right)}
 \dfrac{1}{\sqrt{L''_{\alpha} \left( \dfrac{d_{\alpha}}{R_0}, \dfrac{d_{\alpha}}{R_0} \right)}}
 \dfrac{s^2_{\alpha}}{R_0^2}
 \sin \left[ \displaystyle \frac{1}{h} p_{\rm F} R_0 L_{\alpha} \left( \frac{d_{\alpha}}{R_0}, \frac{d_{\alpha}}{R_0} \right) \right]\Theta(r-d_{\alpha}),
\end{equation}

\begin{equation}
\label{delta_v_nonosc}
 \Delta \varphi_{\rm nonoscil} = \frac{3\sqrt{8}}{4\pi}\frac{e\sqrt{N}}{\varepsilon_0r}
 \sum_{\alpha}\frac{\int\limits_{s_{\alpha}}^r F_{\alpha}\left(\dfrac{r}{R_0},\dfrac{r}{R_0} \right) \dfrac{r^2}{R_0^2}d\dfrac{r}{R_0}}{L_{\alpha}}
 \times\sin\left(\frac{1}{h} p_{\rm F} R_0 L_{\alpha} \right).
\end{equation}
\end{widetext}
Both in~\eqref{delta_v_osc}~and~\eqref{delta_v_nonosc} the summation is carried out over all trajectories $\alpha$ for which $s_{\alpha}<r$.

The resulting expression for $\varphi(r)$ has a spatial scale of the inhomogeneity of the order of the size of the spherical well. For sufficiently large $N$, the potential is determined by the expression~\eqref{delta_v_nonosc}. This result qualitatively differs from the result previously obtained in~\cite{AFM2}.

The structure of the sum in~\eqref{delta_v_nonosc} is such that the potential has extremal points $s_{\alpha}$. For trajectories of type $rot(n, m)$ for small $n, m$ $s_{\alpha}\cong R_0/4,R_0/2,\sqrt{2}R_0/2$ (see Appendix~\ref{app_trajectories}).

\subsection{Numerical calculations}
In this section the theoretical results are confirmed by a numerical summation of exact solutions of the electron wave functions in an infinite potential well.
To simulate the small plasma bubbles with the electron number that is up to $ N = 10 ^ 8 $ let us consider a very simple model.

\begin{figure}[ht]
 \begin{center}
  \includegraphics[width=\linewidth]{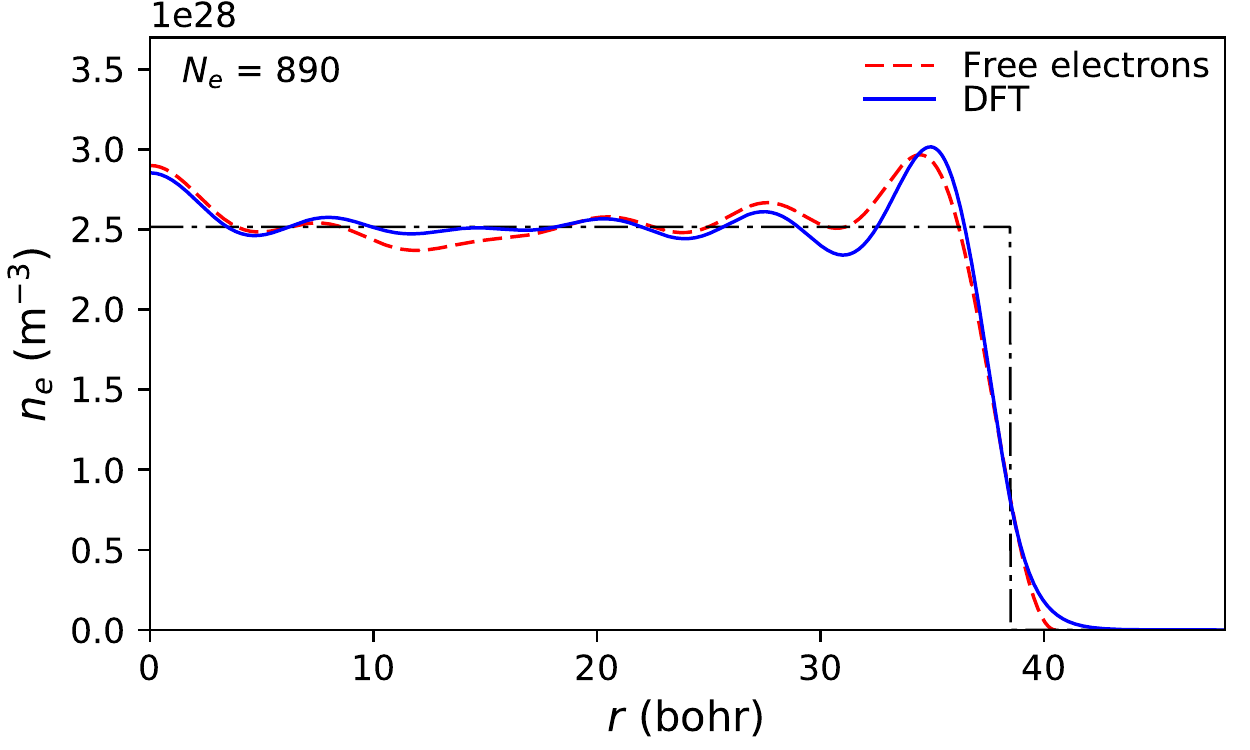}
  \includegraphics[width=\linewidth]{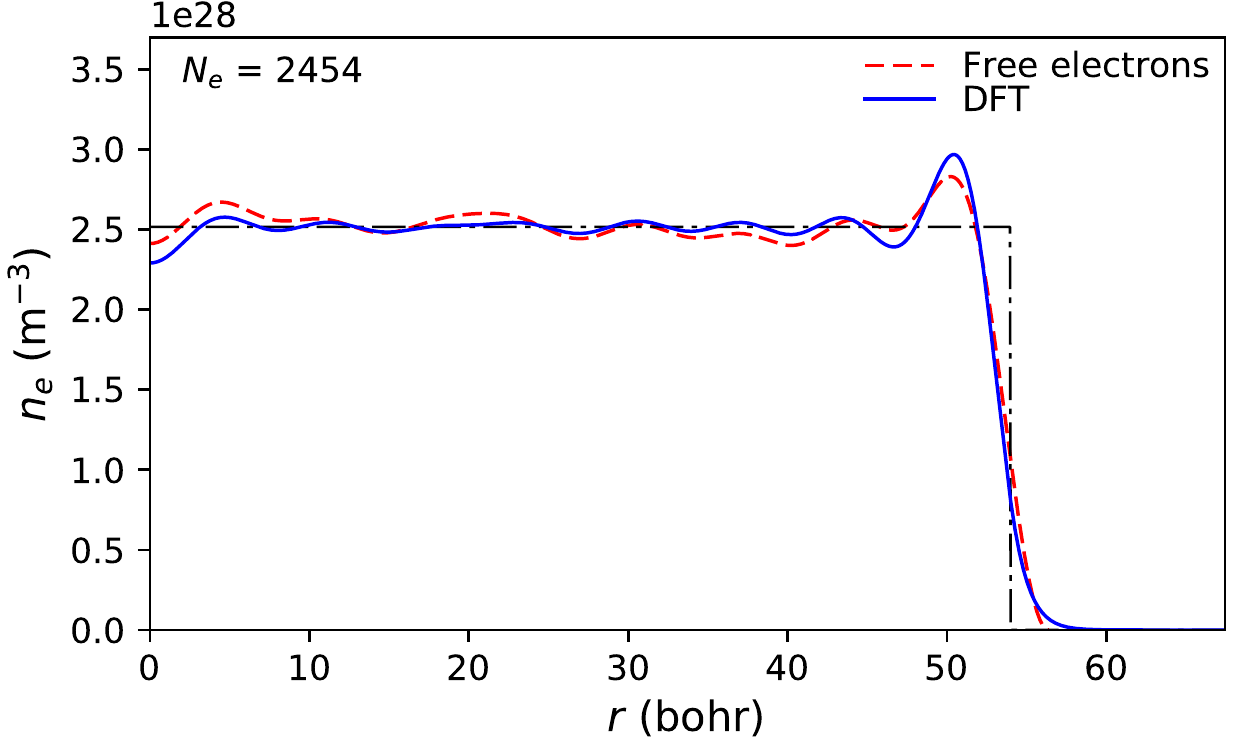}
  \caption{\label{fig:free_vs_DFT} The electron density in the free electrons model (dashed line) and in the DFT (solid line) for 890 (top) and 2454 (bottom) electrons.}
 \end{center}
\end{figure}

We investigate the distribution of $ N $ electrons in the ground
state in a spherically symmetric potential well with impermeable, i.e., infinitely
high, walls. This problem has the complete analytical solution (see, for
example, books by Fluegge and Messiah~\cite{Fluegge, Messiah}).
For the spherical well of radius $R$ the electron number density is
\begin{multline}
\label{rhoreNorm}
\rho_e(r)=\frac{2}{4\pi R^3} \sum_{n_rl} 2(2l+1) j_{l}^2\left(\frac{x_{n_r,l} r}{R}\right) \\
   \times \left(j^2_l(x_{n_r,l})-j_{l-1}(x_{n_r,l})j_{l+1}(x_{n_r,l})\right)^{-1} ,
\end{multline}
where $j_l(x)$ for integer $l$ is a spherical Bessel function of order $l$
and $ x_{n_r,l}$ is its root, which coincides with the root of Bessel function
$J_{l+1/2}(x)$.

The transition to a large number of $ N \gtrsim 10 ^ 8 $ electrons in this
problem allows one to trace in detail the oscillating behavior of the electron
density in the ground state, to see the difference from the quasi-classical
Thomas-Fermi solution and to estimate the magnitude of the electric field, which
can arise during adiabatic compression of a cavity containing a degenerate plasma
with a similarly large number of electrons. The adiabaticity will be understood here
in the spirit of P. Paradoxov~\cite{Paradoksov:1966}.

\begin{figure}[ht]
 \centering
  \includegraphics[width=\linewidth]{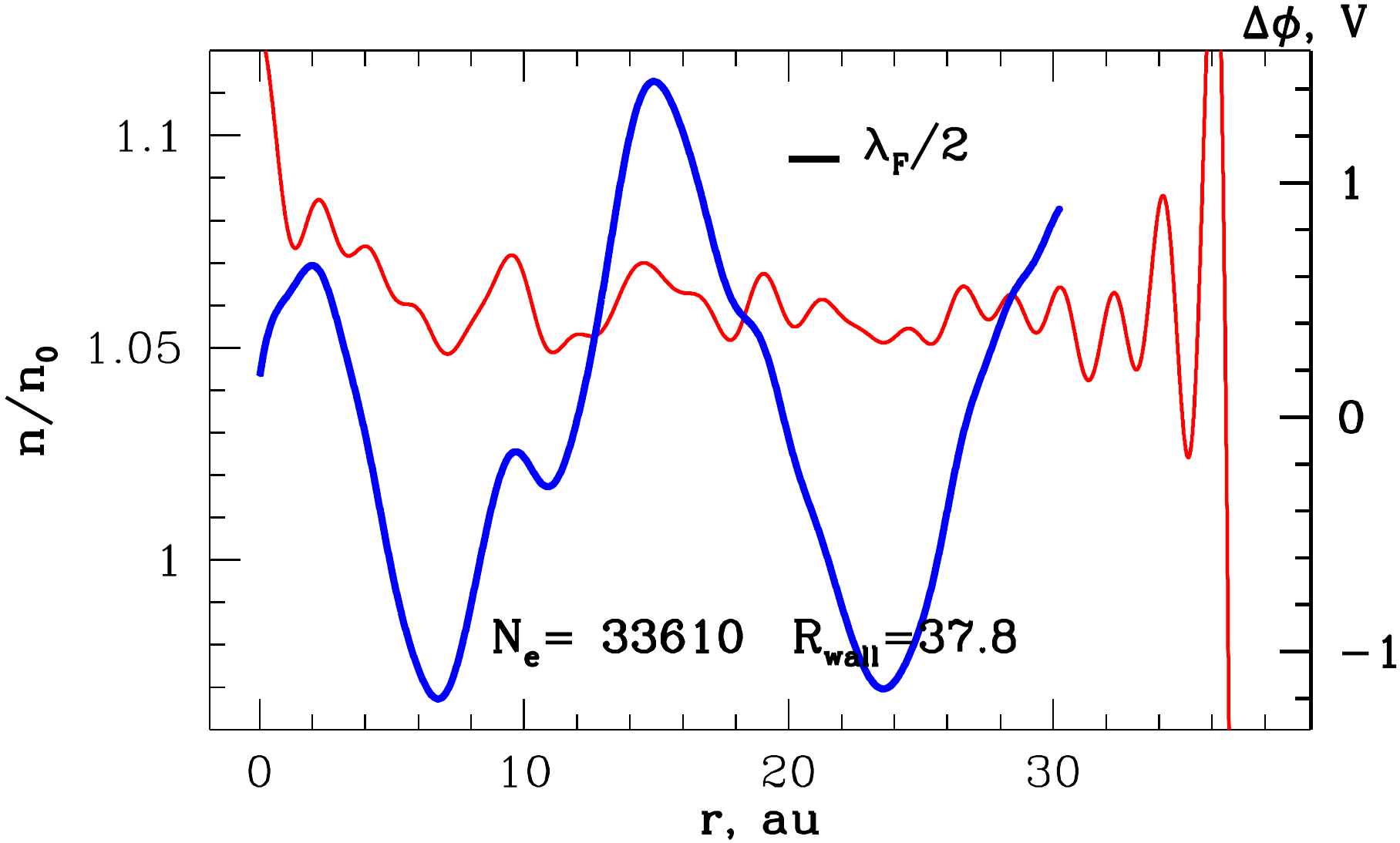}
  \includegraphics[width=\linewidth]{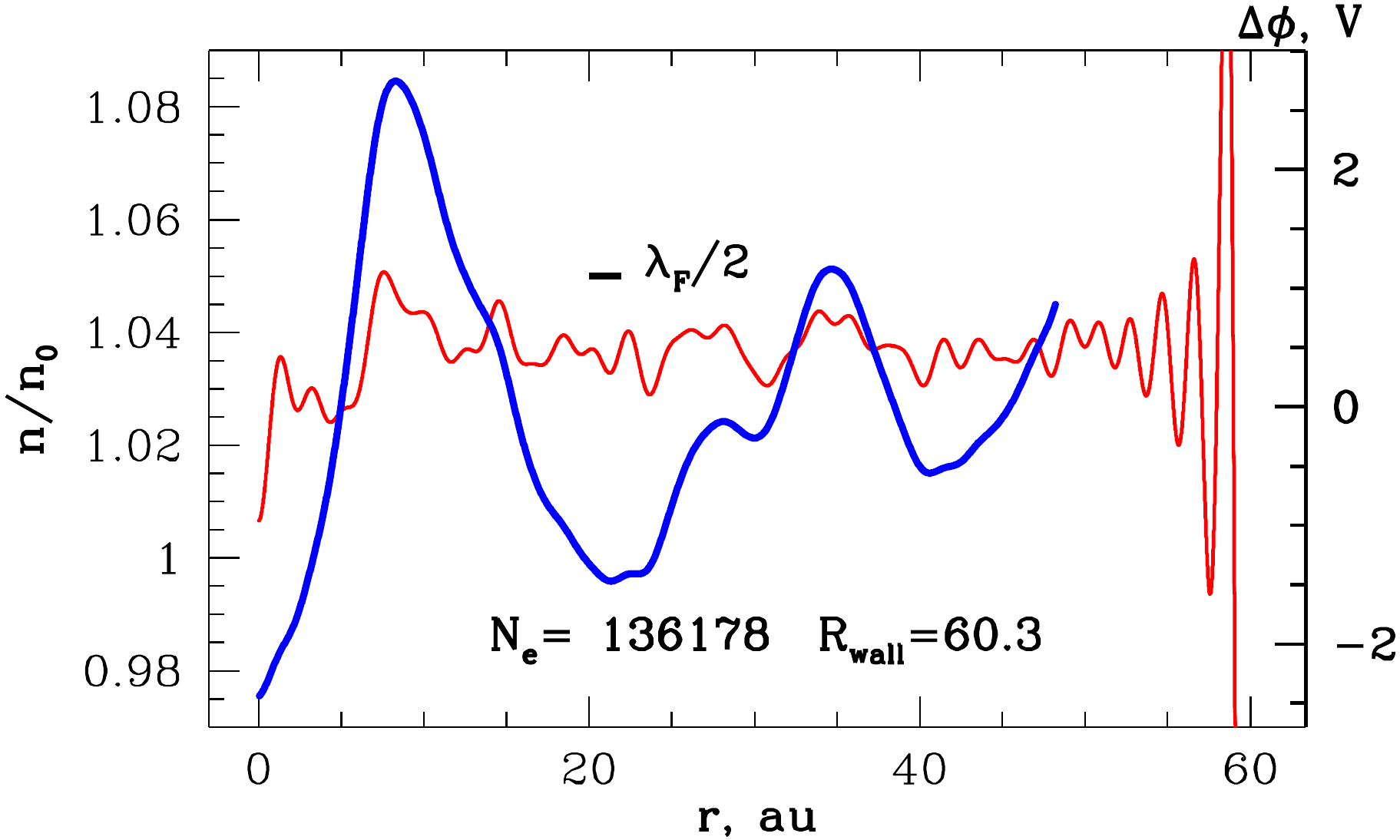}
  \includegraphics[width=\linewidth]{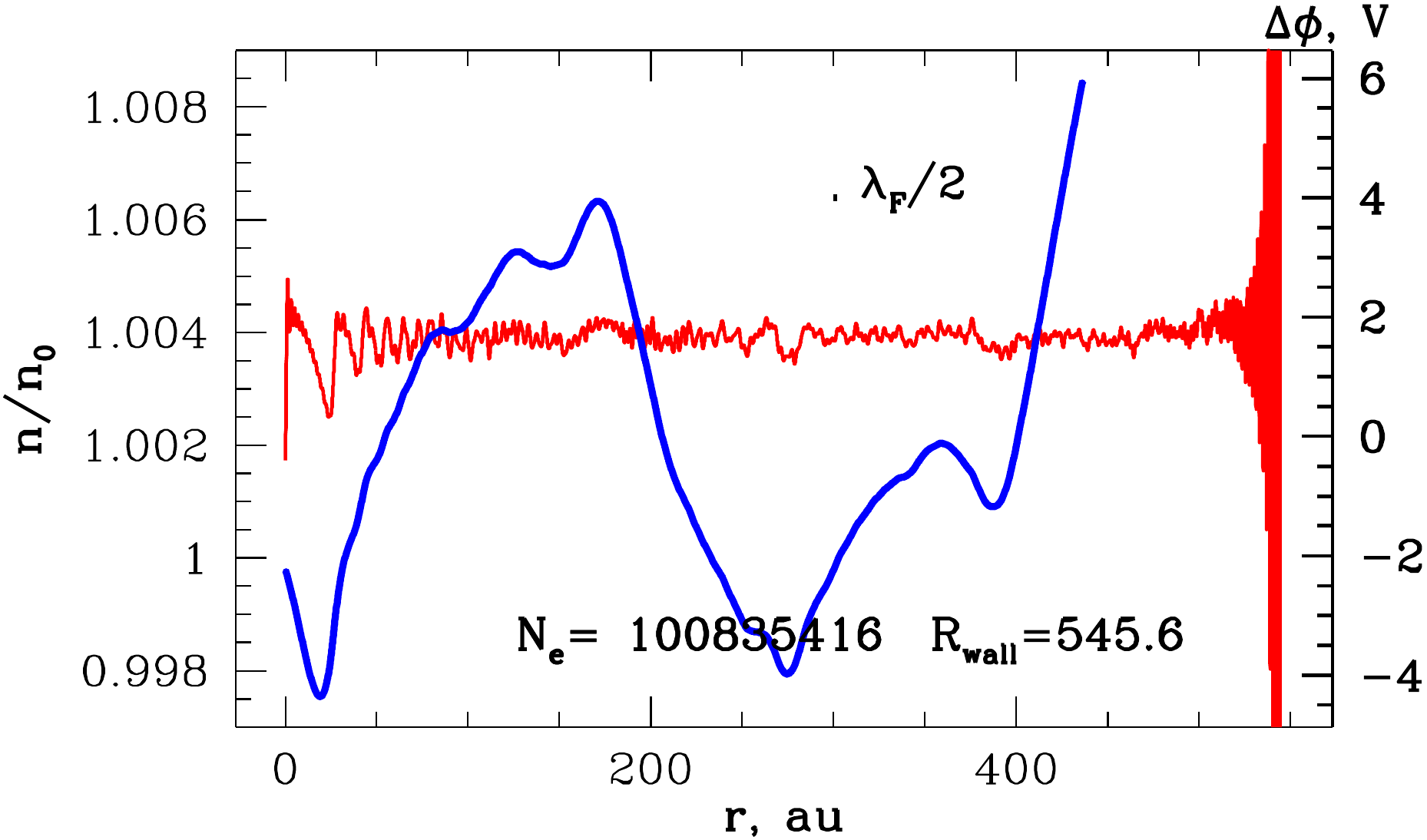}
 \caption{\label{fig:free_osc} Oscillations of the electron density (thin line) and the electrostatic potential (heavy line) for three different numbers of electrons.}
\end{figure}

The results for $\rho_e$ for some values of $N$ are shown in Fig.~\ref{fig:free_vs_DFT} in comparison with DFT results.
If the density $n_i$ is not very high, the behavior of the oscillations in $\rho_e$ and their amplitudes are surprisingly close to the DFT results~\cite{Shidlovski}.
The comparison of DFT predictions for electrostatic potential with this
simple model can be done only if special conditions are imposed.
DFT in jellium model assumes strictly uniform density of ions.
The simple model for the same assumption will predict much larger electric
field, because it lacks correlation and exchange contributions to the potential.

\begin{figure}[ht]
 \begin{center}
  \includegraphics[width=\linewidth]{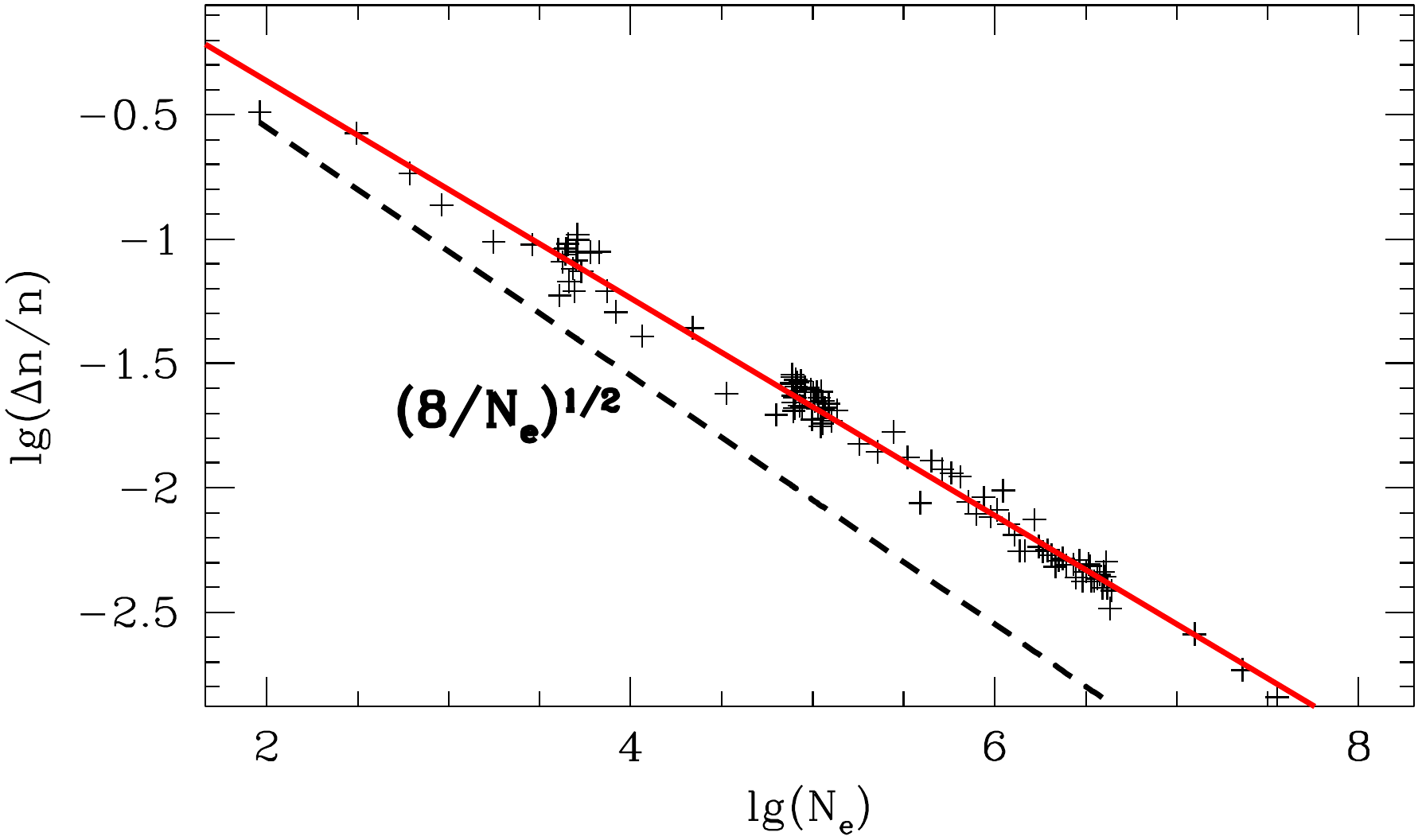}
  \includegraphics[width=\linewidth]{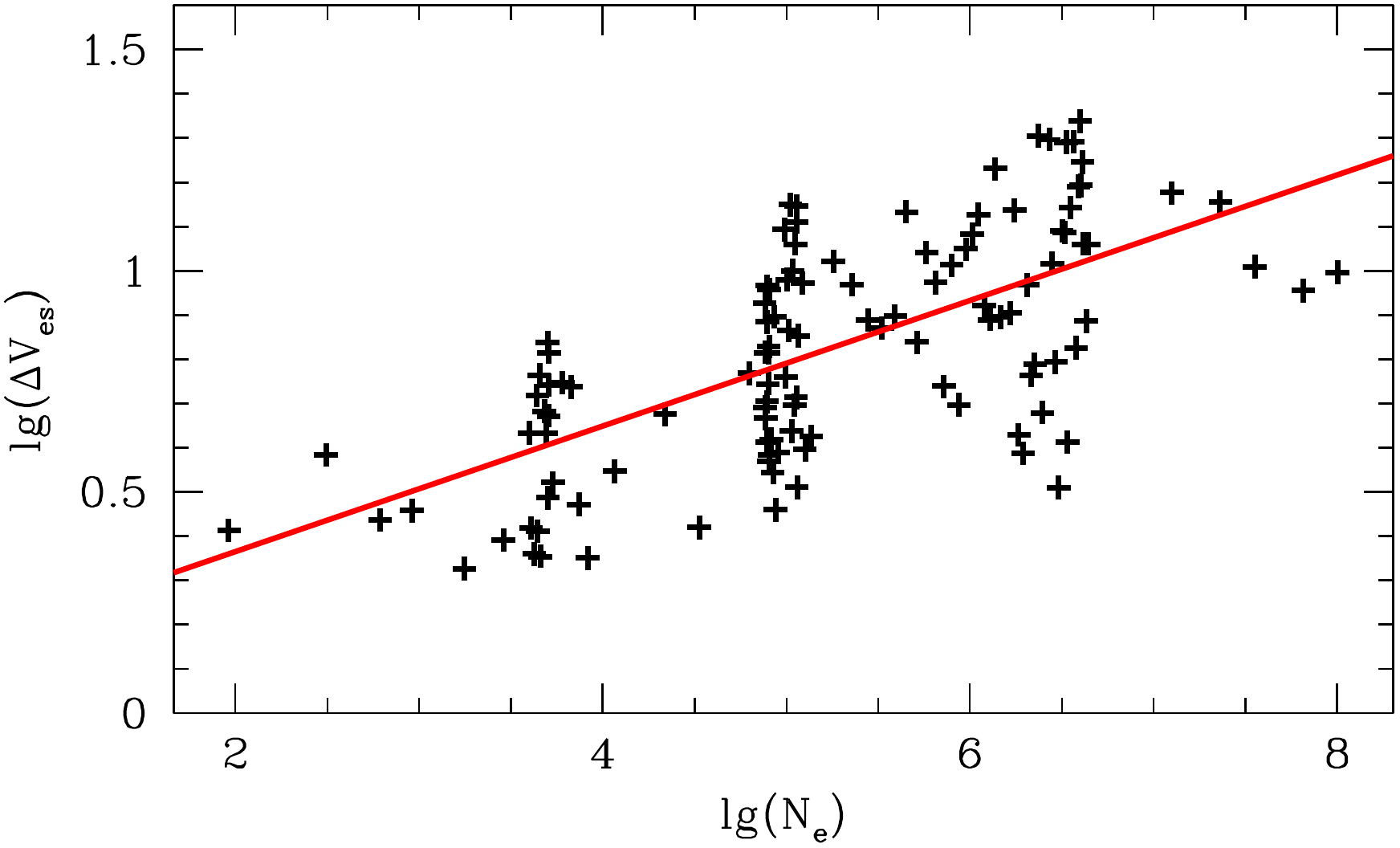}
  \caption{\label{fig:sumDV} Magnitudes of electron density oscillations (top) and potential oscillation depending on number of electrons $N$ at the same density $10^{30} {\rm m}^{-3}$; fitting lines for density and potential correspond to $\Delta n/n \propto N^{-0.437 \pm 0.005}$ and $\Delta V_{\rm ES} \propto n^{1/3}N^{0.142 \pm 0.015}$, respectively.}
 \end{center}
\end{figure}

We get a realistic comparison of electrostatic potentials if we subtract
from the electrostatic potential of $N$ electrons a parabolic least squares
fit, which mimics the electrostatic contribution of ions.
After some numerical experiments we found that good fits are obtained if we
exclude 10\% of inner parts and 20\% of outer parts of the spherical
potential well from the least squares procedure, since the oscillations of
the electron density are especially high there in comparison with DFT jellium model.
The results of computation of the electrostatic potential behavior are shown in Fig.~\ref{fig:free_osc},~\ref{fig:sumDV}. These results confirm the existense of the nonuniformity in the system and the appearance of the spatial scale, which is order of the system size and much larger than the Fermi length. The theoretical results~\eqref{delta_n_sph}, \eqref{delta_v_nonosc} are in a good agreement with numerical results in Fig.~\ref{fig:sumDV}.

\section{Inhomogeneous distribution of the electrons in the compressed gas bubble: DFT calculations}

Previously, we have considered the system of free electrons without interaction and we have shown that its potential 
distribution has an inhomogeneity with a spatial scale of the system size. 
Next we will demonstrate that this effect also takes place in a system of electrons with interaction, such as the compressed bubble of ionized gas.

The numerical simulations of the electron and potential spatial distribution were carried out
applying density functional theory (DFT) in the spherical jellium background model (SJBM).
In the jellium model ions are represented as a continuous fixed distribution of a positive charge.
In this form the spherically symmetric distribution of ions and their potential is possible.
The symmetry allows us to reduce the problem to the one-dimensional formulation.
Here we use the jellium model modification, which is the so-called stabilized jellium model~\cite{PerdewSJ, PerdewSJ2, Kiejna}
with a correction that takes into account an average difference between the jellium and the point ions.

\subsection{Computational method}
For the spherically symmetrical calculations in the jellium model we solve one-dimensional Kohn-Sham equations~\cite{Kohn-Sham}.
When an external potential has a spherical symmetry, the one-electron wavefuction can be decomposed into a radial and a spherical wavefunction.
For the radial wavefunction Kohn-Sham equation has the following form (to simplify the expressions, we use atomic units, where $\hbar=c=e=1$):
\begin{equation}
  \label{Kohn-Sham}
  \left(-\frac{1}{2}\frac{{\rm d}^2}{{\rm d}r^2}+V_{\rm KS}(r)+\frac{l(l+1)}{2r^2}\right)P_{nl}(r)=\varepsilon_{nl}P_{nl}(r),
\end{equation}
\begin{multline}
 V_{\rm KS}(r)=V_{\rm ion}(r)+V_{\rm H}(n_e,r)+\\
 +V_{xc}(n_e,r)+\langle\delta v\rangle_{WS}(n_{\rm ion}(r)),
\end{multline}
where $V_{\rm KS}$ is an effective Kohn-Sham potential, $V_{\rm H}$ and $V_{\rm ion}$ are electrostatic potentials of electrons (Hartree potential) and ionic jellium, respectively, $V_{xc}$ is an exchange-correlation potential, $\langle\delta v\rangle_{WS}$ is a stabilized jellium correction.
We need to emphasize that $V_{\rm H}$ and $V_{\rm ion}$ are usual electrostatic potentials of electrons and ions multiplied by the electron charge $-e$, i.e., $V_{\rm H}=-e\varphi_e=-\varphi_e$ and $V_{\rm ion}=-e\varphi_{\rm ion}=-\varphi_{\rm ion}$.
The electron density is determined by expression
\begin{equation}
 \label{density_sum}
 n_e(r)=2\sum_{n,l}(2l+1)\Theta_{nl}\frac{P^2_{nl}(r)}{4\pi r^2},
\end{equation}
where $\Theta_{nl}=\Theta(E_F-\varepsilon_{nl})$ is the step function.
In the Appendix~\ref{app_jellium} there are more details about potentials and the stabilized jellium model.
Further, when we talk about the electrostatic potential in the jellium model, we mean the sum:
\begin{equation}
 V_{\rm ES}=V_{\rm ion}+V_{\rm H}+\langle\delta v\rangle_{WS}.
\end{equation}

\begin{figure}[ht]
  \begin{center}
   \includegraphics[width=\linewidth]{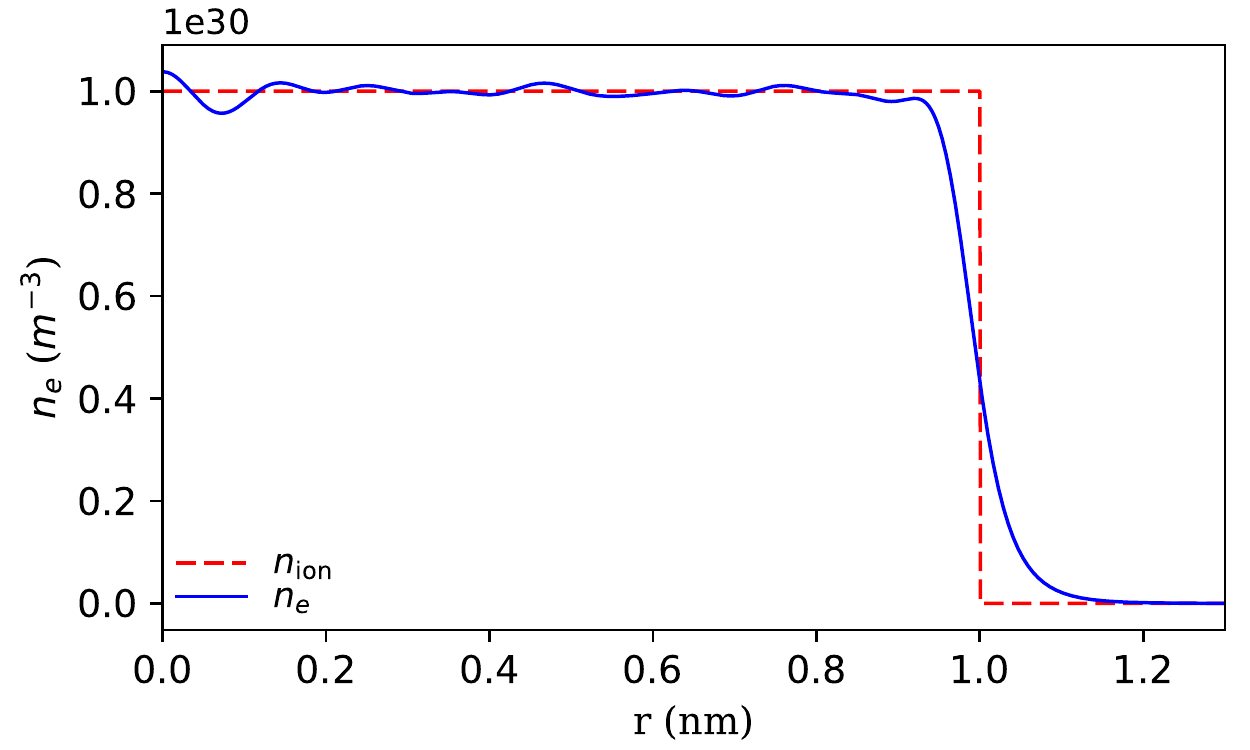}
   \includegraphics[width=\linewidth]{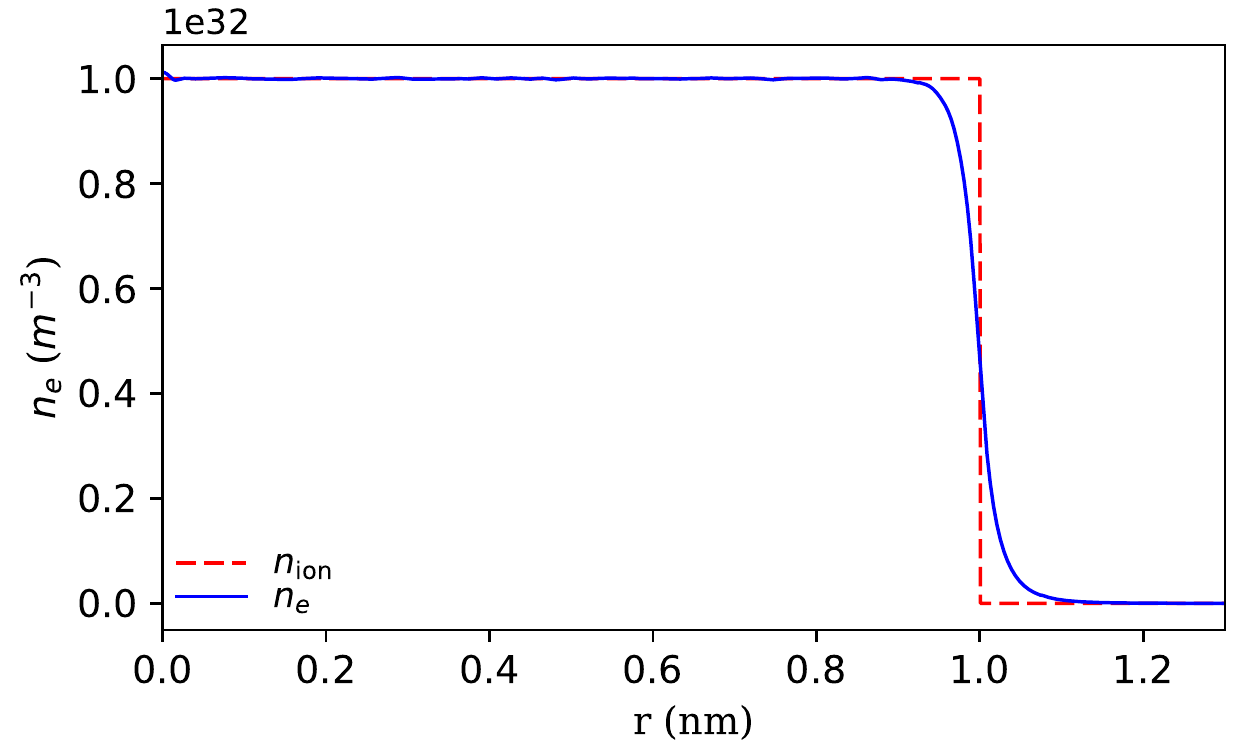}
   \caption{\label{fig:dens_profiles_compare} The electron density (solid line) in a bubble of radius 1 nm at ionic jellium density (dashed line) $10^{30} {\rm m}^{-3}$ with 4188 electrons (top) and at $10^{32} {\rm m}^{-3}$ with 418879 electrons (bottom).}
  \end{center}
\end{figure}

The equations~\eqref{Kohn-Sham} are solved self-consistently using simple iteration method with mixing.
The electron density distribution on the next iteration step is a composition of the previous distribution and $n_e(r)$ obtained from~\eqref{Kohn-Sham} and~\eqref{density_sum}, in which potentials are calculated using the previous electron density distribution.
The iterations continue till the electron density become self-consistent. In all DFT calculation we used PZ (Perdew, Zunger)~\cite{lda_pz} exchange-correlation functional in a local density approximation (LDA) for a spin-unpolarized case to calculate $V_{\rm xc}$.

\subsection{Results of calculations for uniform jellium}

Applying the jellium model we performed a lot of calculations for bubbles of hydrogen with various
sizes and densities of ion jellium.
Only the case of uniform ion density is considered in this section.
In the jellium model it is presented like this: the ion jellium density, which is the same to the positive charge density, equals the average density inside the sphere $\bar n_e=N(3/4\pi R^3)$ and equals zero outside.

The density of ion jellium in performed calculations varies from $10^{29} {\rm m}^{-3}$ to $10^{32} {\rm m}^{-3}$, the number of electrons (the same as number of ions) is up to $4\times10^5$.
Figure~\ref{fig:dens_profiles_compare} shows examples of the electron density spatial distribution in \
the bubble of radius 1 nm for the jellium densities $10^{30} {\rm m}^{-3}$ and $10^{32} {\rm m}^{-3}$. As we can see, the fluctuations of the electron density are much smaller in the second case, in which the number of electrons is hundred times larger.

\begin{figure}[t]
  \begin{center}
   \includegraphics[width=\linewidth]{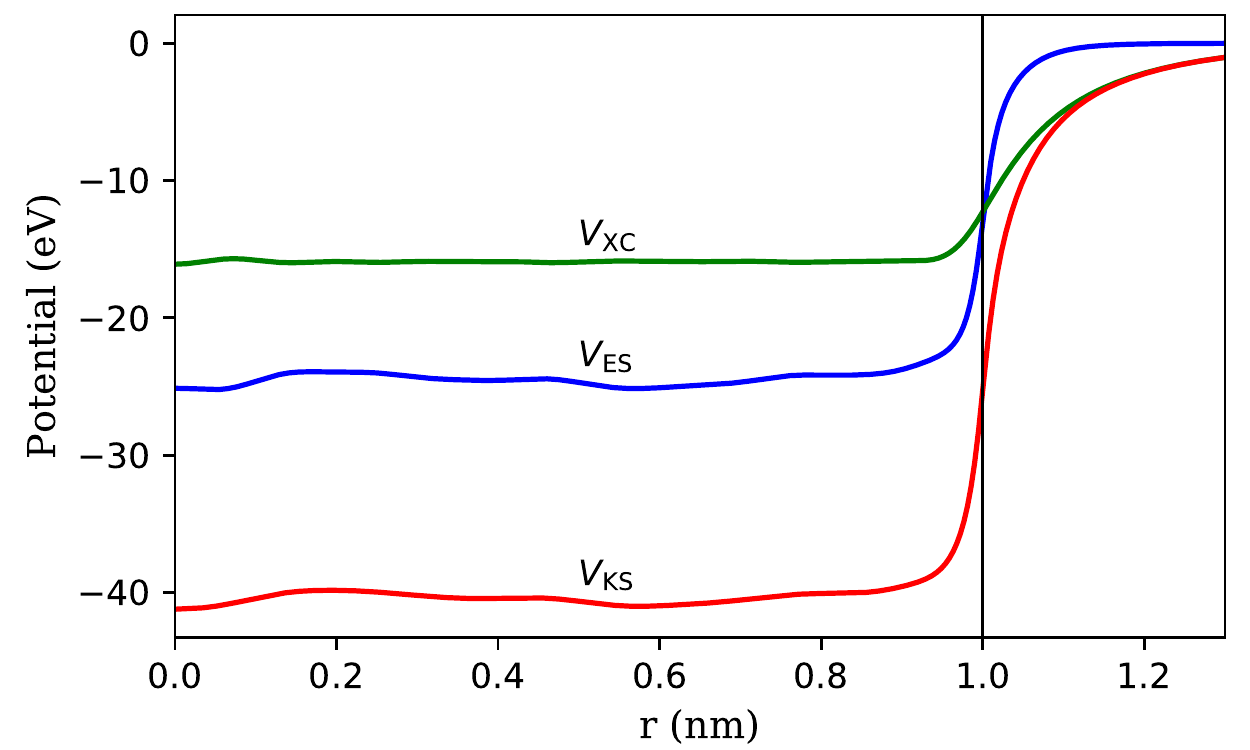}
   \includegraphics[width=\linewidth]{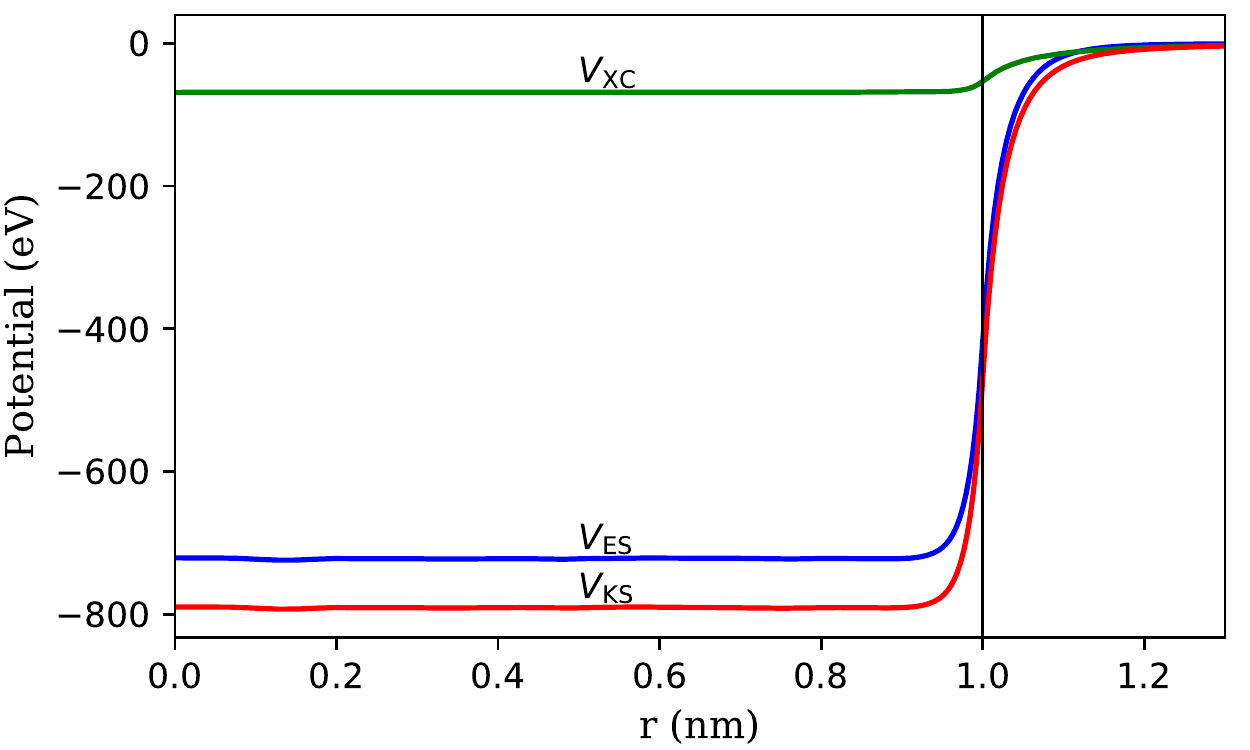}
   \caption{\label{fig:ves_profiles_compare} Potentials in the bubble of radius 1 nm at ionic jellium density $10^{30} {\rm m}^{-3}$ with 4188 electrons (top) and at $10^{32} {\rm m}^{-3}$ with 418879 electrons (bottom).}
  \end{center}
\end{figure}

The potential profiles also become smoother with the growth of the electron number.
Figure~\ref{fig:ves_profiles_compare} shows profiles of the electrostatic potential together with the exchange-correlation and the effective Kohn-Sham potential for the bubbles mentioned above.

 \begin{figure}[ht]
   \begin{center}
   \includegraphics[width=\linewidth]{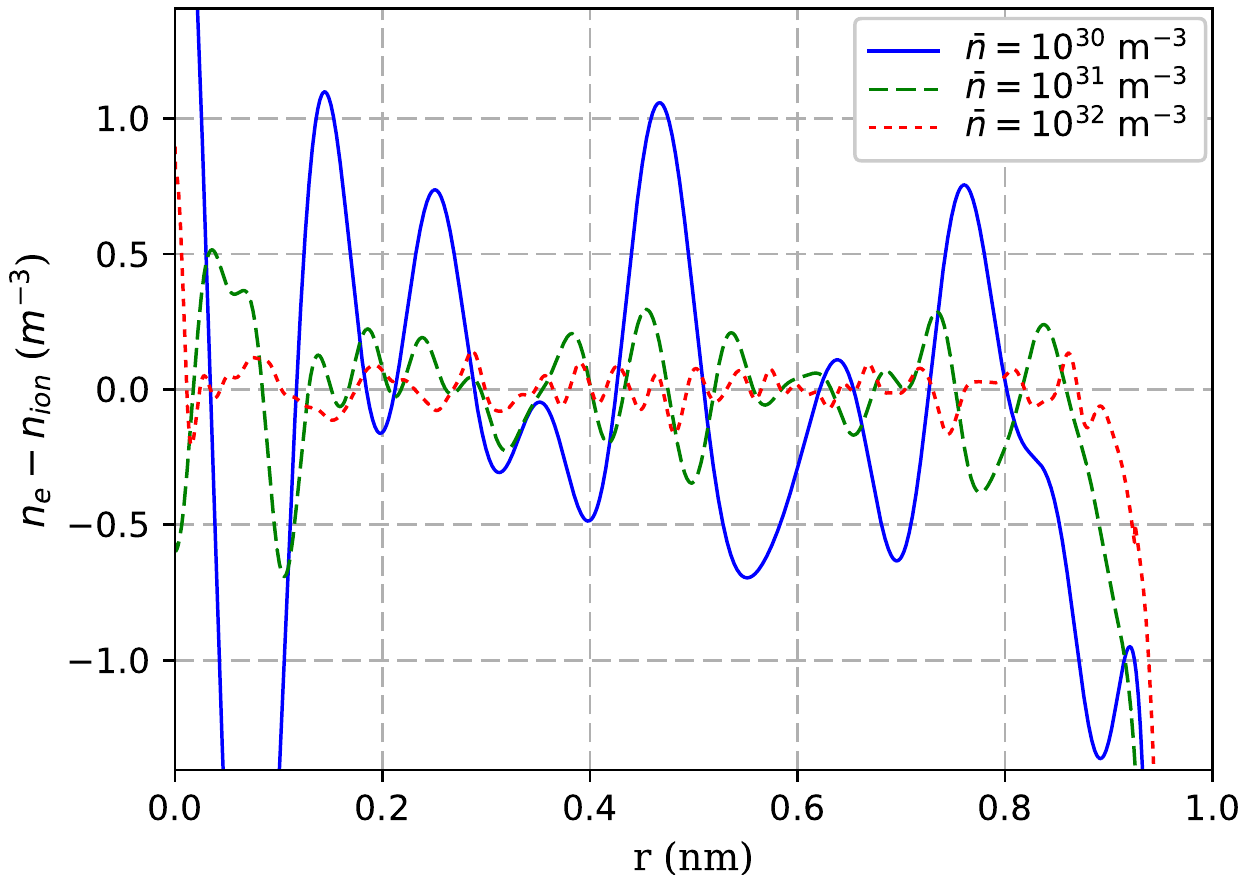}
   \caption{\label{fig:dens_osc_1nm} The electron density oscillations in the bubble of radius 1 nm for three different values of jellium density: $10^{30}, 10^{31}$ and $10^{32} {\rm m}^{-3}$.
   }
   \end{center}
 \end{figure}

 \begin{figure}[ht]
   \begin{center}
   \includegraphics[width=\linewidth]{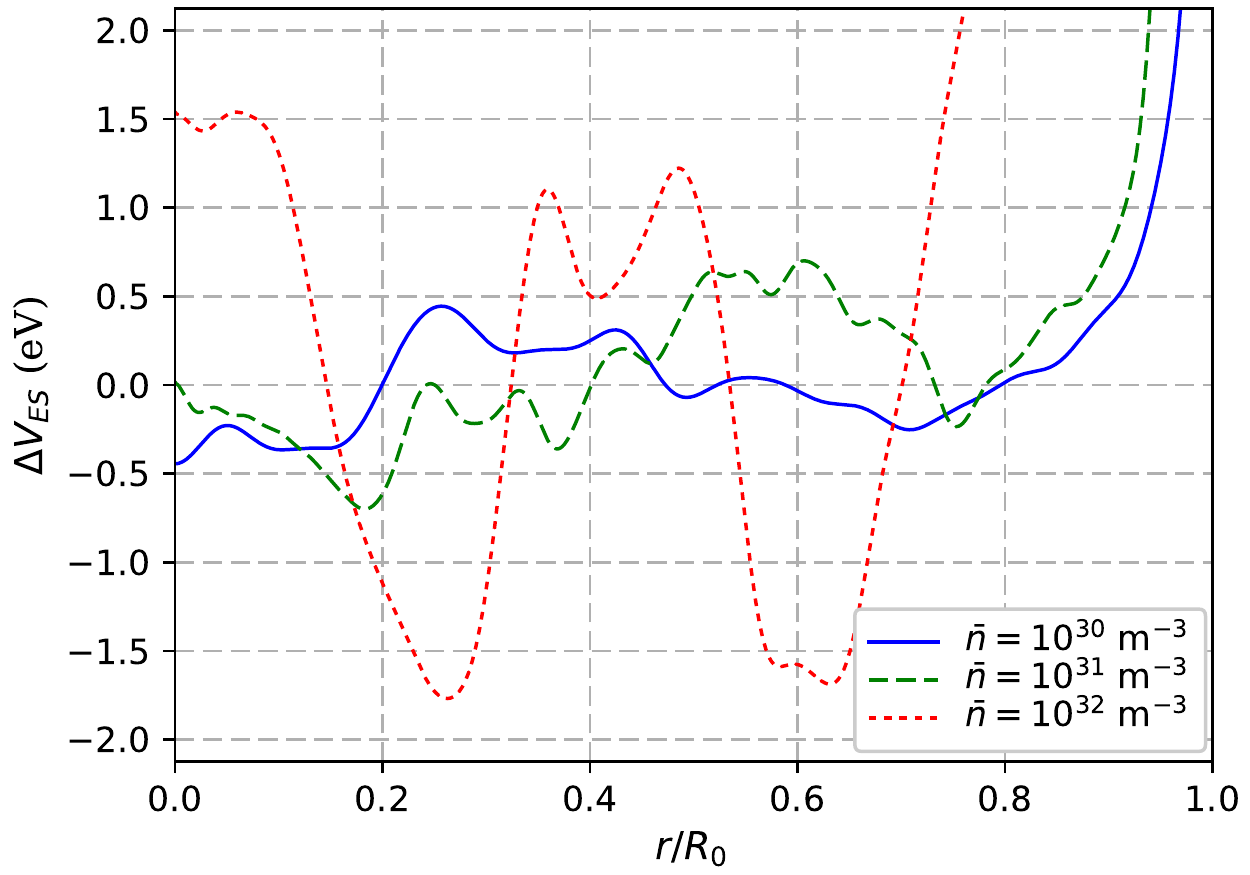}
   \caption{\label{fig:ves_osc_1nm} The electrostatic potential oscillations in the bubbles at densities $10^{30}, 10^{31}$ and $ 10^{32} {\rm m}^{-3}$ and radii 1.34~nm, 1.5~nm and 0.5~nm, respectively.
   }
   \end{center}
 \end{figure}

Next let us discuss in detail the oscillations of electron density and electrostatic potential.
We can simply take a difference between the electron density and the ion jellium density $n_e-n_{\rm ion}$ as oscillations of the electron density, since the ion jellium density is uniform and approximately equals the average electron density.
We take half of a difference between maximum and minimum values of $n_e-n_{\rm ion}$
as a magnitude of the electron density oscillations $\Delta n$
in the intermediate region, excluding part near the border and the central part of a bubble,
since the oscillations are always very high at the center.
These oscillations in the bubble of radius 1~nm at three different densities of ion jellium are presented in Fig.~\ref{fig:dens_osc_1nm}. It is clearly visible that the amplitudes of the electron density oscillations decrease with increasing number of electrons, as should be according to~\eqref{delta_n},
\begin{equation}
\label{delta_n_au}
 \dfrac{\Delta n}{n} \approx \left(\dfrac{8}{N_e}\right)^{1/2}.
\end{equation}

To extract oscillations of electrostatic potential $\Delta V_{ES}$ we can, for example, subtract an average value from $V_{ES}$, which is approximately expressed through the Fermi energy $E_F$ ($\bar{n}$ is an average density):
\begin{equation}
 \label{deltaVes}
 \Delta V_{ES}=V_{ES}+V_{\rm xc}(\bar{n})+(E_F(\bar{n})-\varepsilon^{max}),
\end{equation}
since
\begin{equation}
 V_{ES}+V_{\rm xc} = V_{KS} \approx -E_F.
\end{equation}

From~\eqref{delta_v_nonosc} we have
\begin{equation}
\label{delta_v_au}
\phi (r) \propto \frac{\sqrt{N_e}}{R_0} \propto n_0^{1/3}N_e^{1/6}.
\end{equation}

 \begin{figure}[ht]
   \begin{center}
   \includegraphics[width=\linewidth]{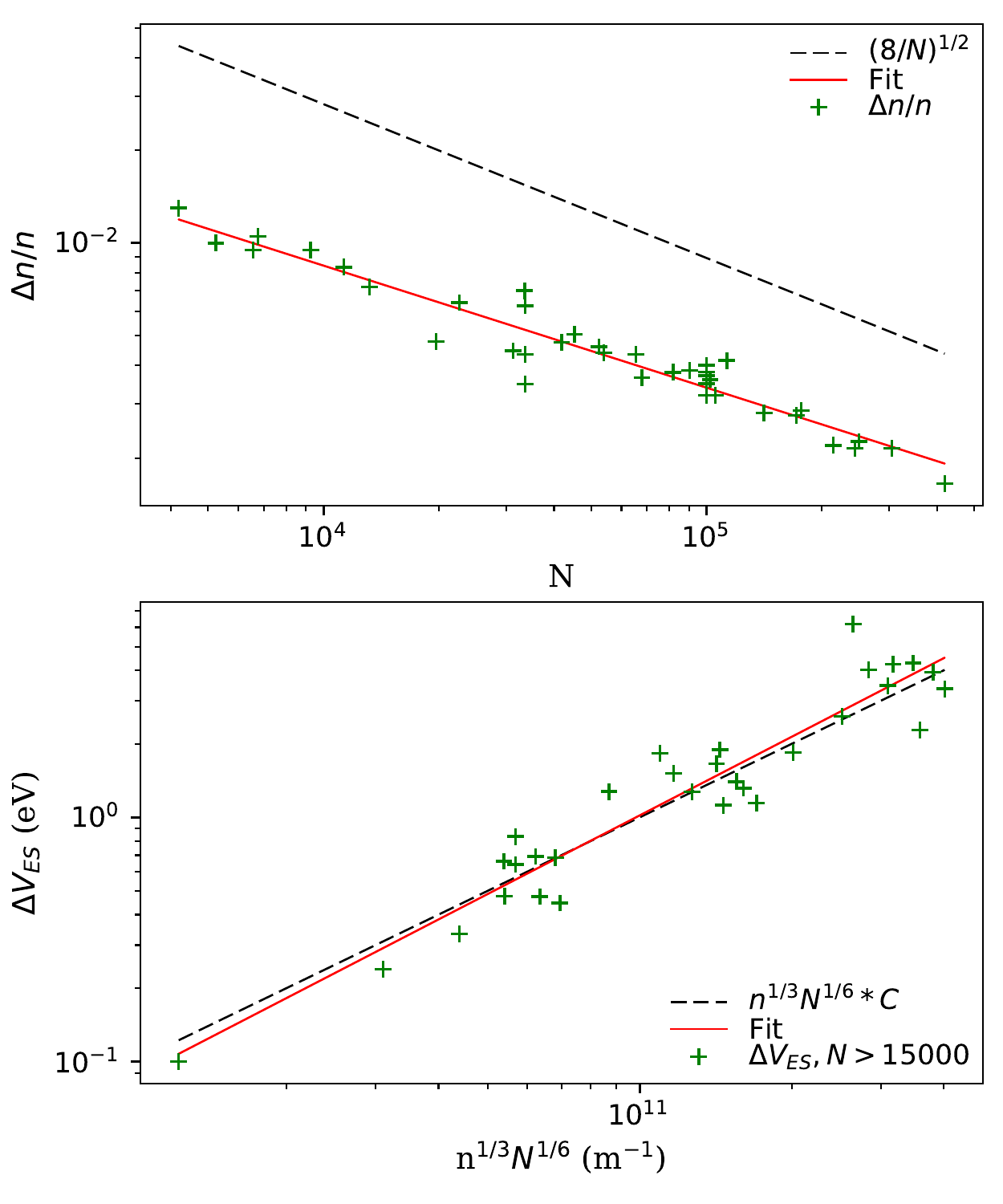}
   \caption{\label{fig:all_results} The dependence of magnitudes of the electron density oscillations on the number of electrons $N$ (top) and potential oscillation depending on value~\eqref{delta_v_au} (bottom) in the logarithmic scale; analytical dependencies~\eqref{delta_n_au} and~\eqref{delta_v_au} are shown by dashed lines.}
   \end{center}
 \end{figure}

All results on oscillations of electron density and potential for a set of calculations are presented in Fig.~\ref{fig:all_results}. The first chart shows the magnitude of electron density oscillations depending on number of electrons in a logarithmic scale. There is also dependence~\eqref{delta_n} and a fit for a power law, which was obtained using the least square method. The fitting line corresponds to law $\Delta n/n\propto N^{-0.4}$.

The magnitude of potential oscillations depending on density is shown on the second chart in Fig.~\ref{fig:all_results} with law~\eqref{delta_v_au} and a power law fit.
 We excluded the results, obtained for $N<15000$ there, because, for a relatively small number of electrons the potential oscillations do not satisfy the law~\eqref{delta_v_au} and they are several times bigger than they should be according to the power law fit.
 The fitting line for the potential oscillations corresponds to dependence $\Delta V_{\rm ES}\propto \varphi^{1.07}$, where $\varphi$ is from~\eqref{delta_v_au}.
 It means that the magnitudes of the potential oscillations are in a good agreement with~\eqref{delta_v_au}.

\section{The manifestation of quantum shell effects in hydrodynamic processes}

\subsection{Hydrostatic equilibrium of the compressed gas bubble}

 \begin{figure}[ht]
   \begin{center}
   \includegraphics[width=\linewidth]{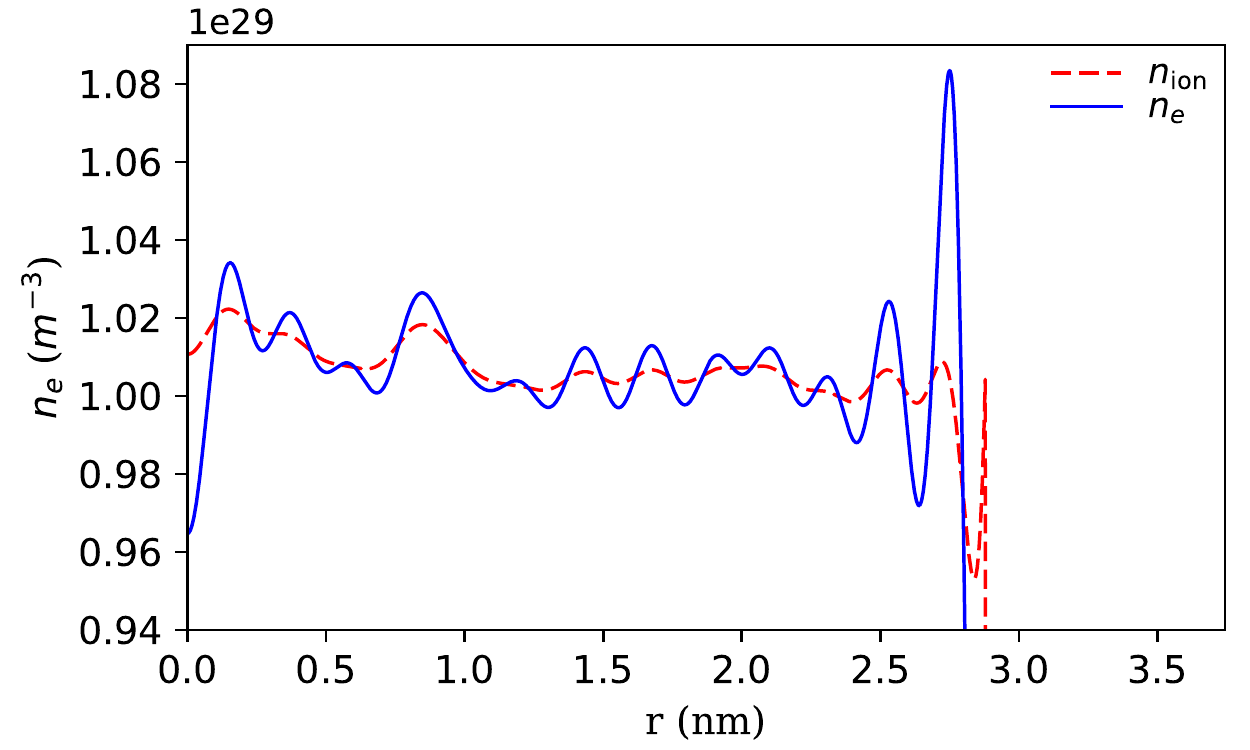}
   \includegraphics[width=\linewidth]{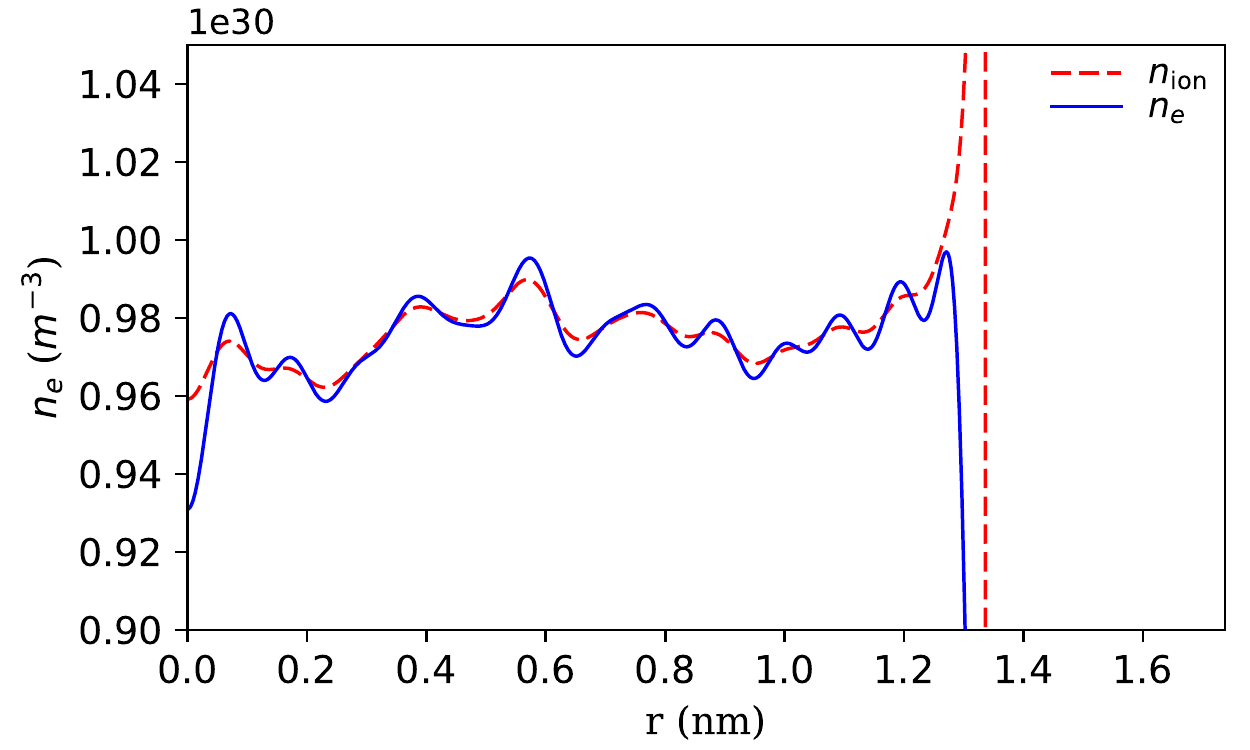}
   \includegraphics[width=\linewidth]{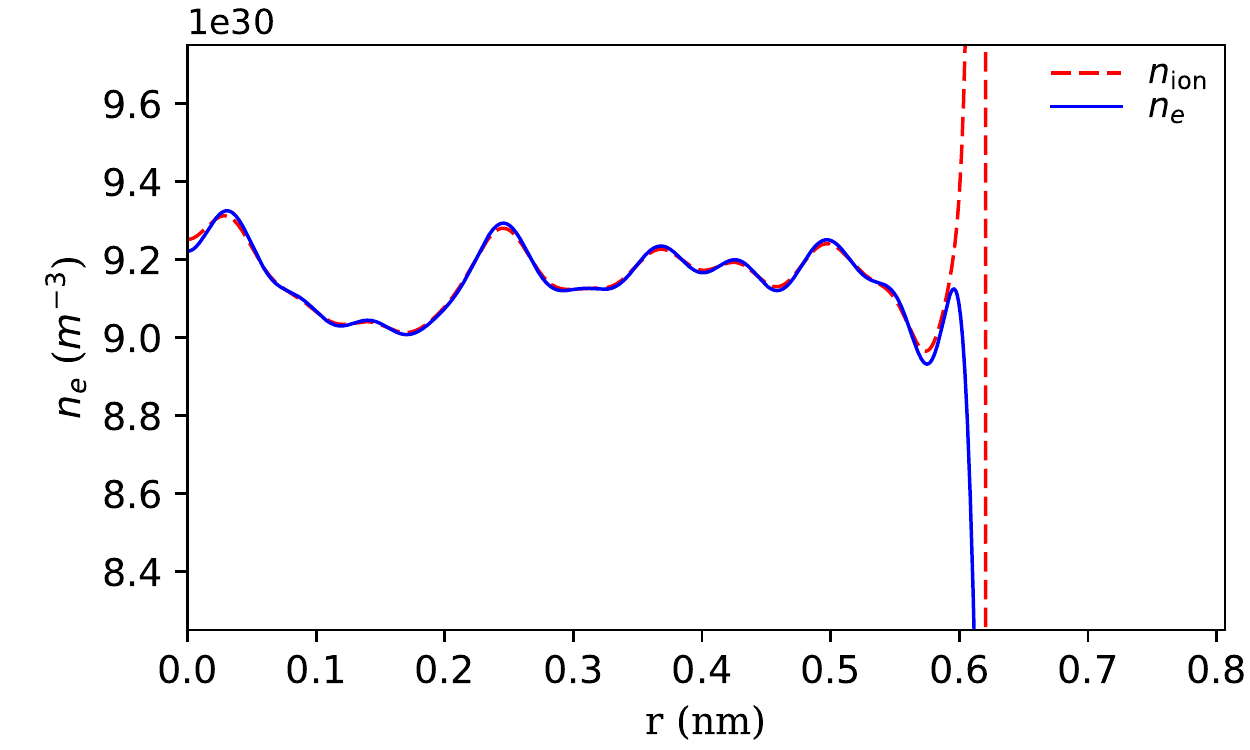}
   \caption{\label{fig:ion_hydro} The electron densities (solid lines) and the densities of ions in equilibrium (dashed lines) at ion temperature 10 eV in the bubble with 10000 electrons and average jellium density $10^{29} {\rm m}^{-3}$ (top), $10^{30} {\rm m}^{-3}$ (middle), and $10^{31} {\rm m}^{-3}$ (bottom).}
   \end{center}
 \end{figure}

Previously, we have considered the case, when the ion jellium distribution is uniform, however, the uniform distribution actually is not equilibrium, and ions will redistribute.
One of the methods to calculate the equilibrium distribution of the ion density is to solve the
hydrostatic equation for ion liquid.
In a statical equilibrium an ion pressure is balanced with an electric field force, i.e. (we use atomic units here),
\begin{equation}
  \label{hydrostatic_ion}
 -\frac{1}{n_{\rm ion}}\nabla P - Z_{\rm ion}\nabla \varphi = 0,
\end{equation}
where $P=n_{\rm ion}kT_{\rm ion}$ is the ion pressure, $Z_{\rm ion}$ is the ion charge, $\varphi$ is the electrostatic potential, which is defined by Poisson equation
\begin{equation}
 \label{Poisson}
 \nabla^2\varphi=4\pi (n_e-Z_{\rm ion}n_{\rm ion}).
\end{equation}
In the case of a spherical symmetry one can use formula~\eqref{V_H_formula} instead of ~\eqref{Poisson}. At constant temperature we can simply integrate Eq.~\eqref{hydrostatic_ion} and obtain
\begin{equation}
  \label{hydrostatic_ion_int}
 \frac{kT_{\rm ion}}{Z_{\rm ion}}\ln(n_{\rm ion})=-\varphi+C.
\end{equation}

As a result, to obtain the electron density distribution with the ions in equilibrium we should self-consistently solve Eq.~\eqref{Kohn-Sham} together with Eq.~\eqref{hydrostatic_ion_int}.
But calculations become more complicated than in case of the uniform ion jellium, so we managed to obtain results for systems with $ \lesssim 10^4$ electrons.

Figure~\ref{fig:ion_hydro} shows the results of calculations for ions in equilibrium at constant ion temperature 10~eV for three density values, in larger scale.
The ions are confined by a potential wall at the bubble boundary to prevent an expansion of matter.
The ions accumulate at the boundary, because of the presence of this wall and the fact that electrons are partially behind the sphere boundary,
thus creating the excess of positive charge inside the bubble.
In Fig.~\ref{fig:ion_hydro} we can see the electron density and the density of ions getting closer to each other with the average density increasing.

\begin{figure}[ht]
   \begin{center}
   \includegraphics[width=\linewidth]{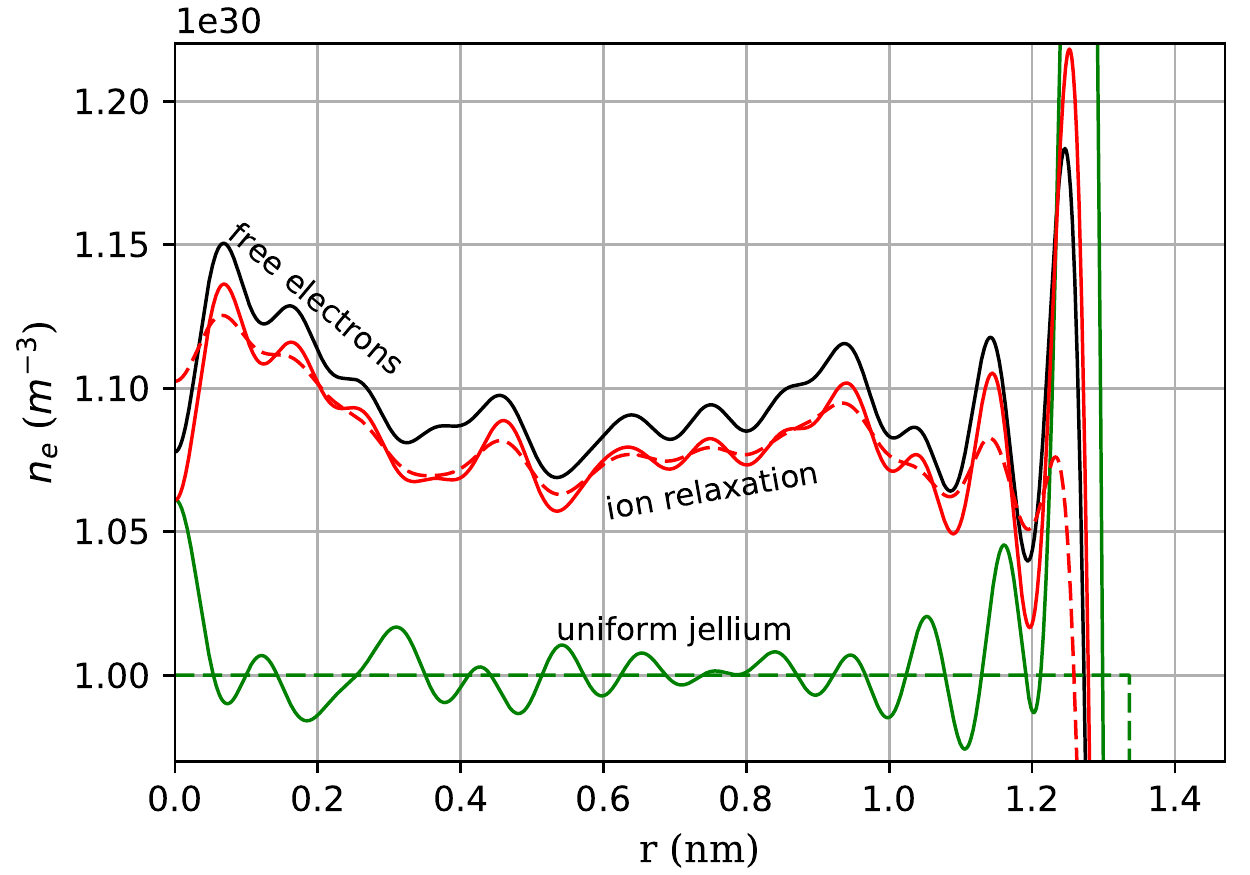}
   \includegraphics[width=\linewidth]{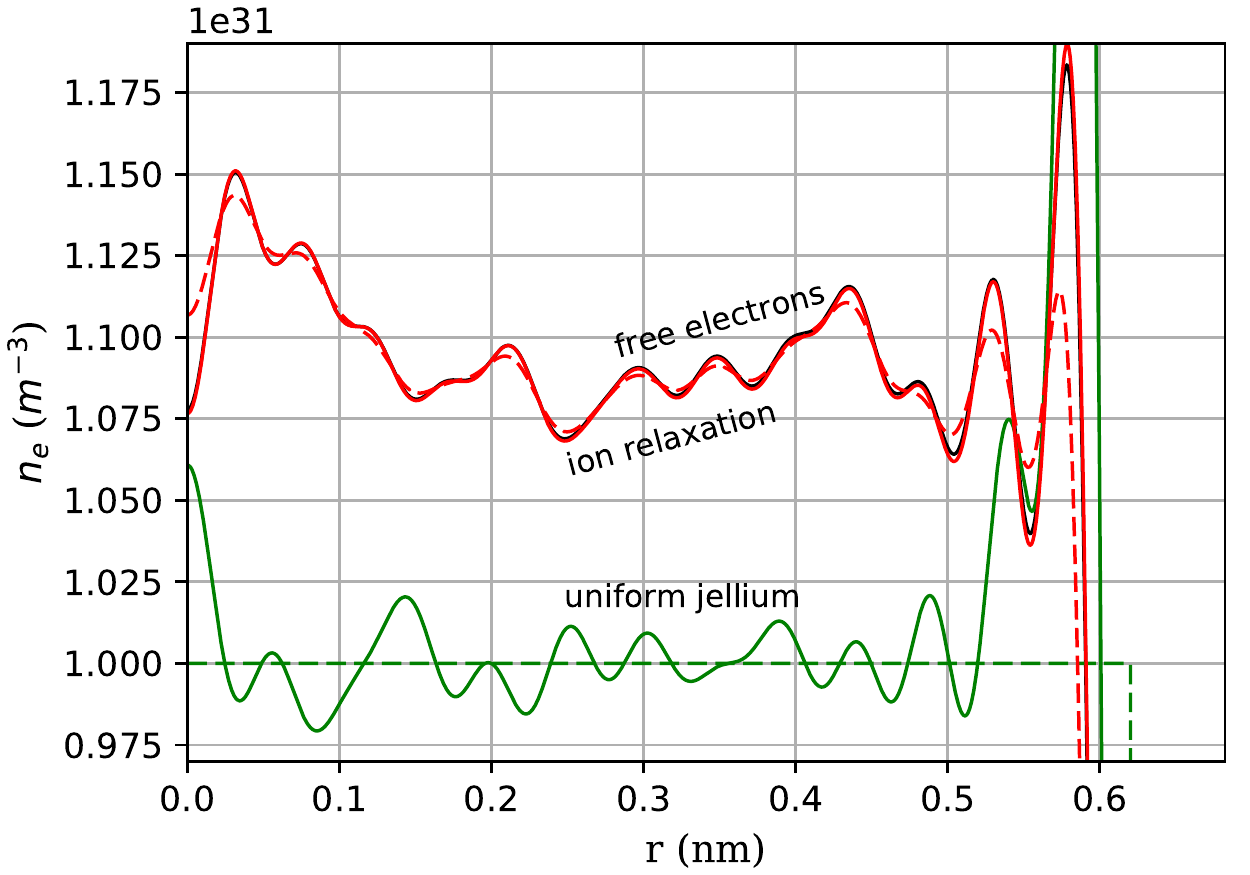}
   \caption{\label{fig:free_uniform_ionhydro_compare} The electron density (solid lines) and the ion density (dashed lines) in different cases: with uniform ionic jellium, with ionic relaxation at temperature 10~eV, and also for free electrons.
   The number of electrons is 10000, average density is $10^{30} {\rm m}^{-3}$ (top) and $10^{31} {\rm m}^{-3}$ (bottom).}
   \end{center}
 \end{figure}

A comparison between the uniform jellium model, the hydrostatic ion relaxation model and the free electrons model is presented in Fig.~\ref{fig:free_uniform_ionhydro_compare}. In the case of ionic relaxation,
the electron density gets very close to the free electron distribution, when the average density grows.
The electrostatic potential oscillations have almost the same form as the density since $n_{\rm ion}\approx n_e$, and from Eq.~\eqref{hydrostatic_ion_int} we can get
\begin{equation}
 -\Delta\varphi+C=
 \dfrac{kT_{\rm ion}}{Z_{\rm ion}}\ln\left(1+\dfrac{\Delta n_{\rm ion}}{\bar{n}}\right)\approx
 \dfrac{kT_{\rm ion}}{Z_{\rm ion}}\dfrac{\Delta n_e}{\bar{n}}.
\end{equation}

Let us consider equation for the quantum electron fluid without Bohm potential~\cite{Shukla} but with
the oscillation potential $U_{\rm osc}$
\begin{equation}
 \label{hydrostatic_electron_osc}
 -\frac{1}{m_{e}n_{e}}\nabla P_e + \frac{e}{m_{e}}\nabla \varphi + \frac{e}{m_{e}}\nabla U_{\rm osc} = 0,
\end{equation}
that should give the electron density similar to one obtained from DFT calculations, which is also the same as free electrons density. Since $\nabla P_e/n_e=\nabla E_F $
\begin{equation}
 eU_{\rm osc} = E_F - e\varphi+{\rm const} \approx E_F(n_e^{\rm free})+{\rm const},
\end{equation}
because $\varphi\longrightarrow0$ at $T_{\rm ion}\longrightarrow0$.

\subsection{Nontrivial hydrodynamic of the compressible gas bubble}
The quantum shell effects are manifested in the hydrodynamic of the compressible gas bubble, which is a mixture of a gas of degenerate electrons and a gas of classical ions. It leads to a new type of cumulation.

As shown in the previous section, the concentration of degenerate electrons changes due to quantum shell effects. Under the action of the arising electric field the ions relax to a concentration $n_i=n_e=n_{\rm free}$, which affords the electroneutrality of the gas bubble. We consider a compression regime for which the relaxation time of ions is much less than the compression time. Therefore the system is locally electroneutral at all times. We show that the compression of the bubble has the nontrivial hydrodynamic behavior.

The system of hydrodynamic equations, which describes the dynamics of electrons and ions, has the following form~\cite{Shukla}:
\begin{gather}
 \frac{\partial n_e}{\partial t} + \nabla(n_ev_e) = 0,\\
 \frac{\partial v_e}{\partial t}+v_e\frac{\partial v_e}{\partial r} = -\frac{1}{\rho_e}\frac{\partial p_e}{\partial r} + \frac{e}{m_e}\nabla(\varphi) - \frac{e}{m_e}\nabla U_e,\\
 \frac{\partial n_i}{\partial t} + \nabla(n_iv_i) = 0,\\
 \frac{\partial v_i}{\partial t}+v_i\frac{\partial v_i}{\partial r} = -\frac{1}{\rho_i}\frac{\partial p_i}{\partial r} - \frac{e}{m_i}\nabla\varphi,\\
 p_e=\frac{(3\pi^2)^{2/3}}{5}\frac{h^2}{m_e}n_e^{5/3}.
\end{gather}
We neglected the Bohm’s term and took into account the quantum orbital effects, as it was proposed in the previous section.
\begin{equation}
 eU_e=\frac{(3\pi^2)^{2/3}}{2}\frac{h^2}{m_e}(n_e^{\rm free})^{2/3}.
\end{equation}
Since we analyze small perturbation of spatial distribution,
\begin{equation}
 \frac{\Delta n}{n_0} \sim \frac{1}{\sqrt{N}},
\end{equation}
we consider the problem in the linear approximation
\begin{equation}
 n_i=n_0+\delta n_i,\quad n_e=n_0+\delta n_e,\quad \delta n_e\approx\delta n_i.
\end{equation}

In the analyzed situation
\begin{equation}
  0 = -\frac{1}{\rho_e}\frac{\partial p_e}{\partial r} + \frac{e}{m_e}\nabla\varphi - \frac{e}{m_e}\nabla U_e.
\end{equation}
Therefore, in the linear approximation, the
system of equations takes the form
\begin{gather}
  \frac{\partial v_i}{\partial t} = -\frac{1}{\rho_i}\frac{\partial p_i}{\partial r} - \frac{1}{\rho_i}\frac{\partial p_e}{\partial r} - \frac{e}{m_i}\nabla U_e,\\
  \frac{\partial (\delta n_i)}{\partial t} + n_0\nabla(v_i) = 0.
\end{gather}

Taking into account $p_i\ll p_e$ for the velocity potential $v_i=-{\rm grad}\psi$, we obtain the equation
\begin{equation}
 \frac{1}{c^2}\psi_{tt}-\Delta\psi = \frac{1}{c^2}\frac{e}{m_i}\frac{\partial U}{\partial t},\quad
 c^2 = \frac{\partial p_e}{\partial \rho}.
\end{equation}
This is the wave equation with the source. The solution of this equation (expressed via so-called retarded potentials)
shows that the presence of the shell quantum effects leads to the generation of sound waves in the volume of the compressible gas bubble.
Because of the non-monotonic spatial distribution of the ion density, sound waves are transformed into shock waves~\cite{Whitham}.
This leads to a nontrivial compression mechanism of such a system, which differs from the known adiabatic compression.

\section{Limiting factors}

In this part we consider several factors, which can lead to damping of oscillations, such as a
different symmetry of the system, the temperature of electrons and non-sphericity of the system
boundary.

\subsection{System symmetry}

As an example of system with different symmetry, we chose a cylinder, which is infinite along its symmetry axis. The method for calculation of electron distribution in an infinite cylinder is presented in the Appendix~\ref{app_cylinder}.
To compare the manifestation of the effect in the spherically symmetrical geometry with
its display in the axial symmetrical geometry, we made several calculations for a sphere and an infinite cylinder at equal values of radius and ionic jellium density.

 \begin{table}[ht]
  \caption{\label{tab:oscillation_amplitude_cylinder} A comparison of the electron density and the electrostatic potential oscillations obtained in the numerical simulations for a spherical symmetry and for a cylinder symmetry.}
  \begin{ruledtabular}
  \begin{tabular*}{\linewidth}{cD{.}{.}{3}D{.}{.}{2}D{.}{.}{2}D{.}{.}{3}D{.}{.}{3}}
     & & \multicolumn{2}{c}{Sphere} & \multicolumn{2}{c}{Cylinder}\\
     $\bar{n}$ & \multicolumn{1}{c}{$R$} &
     \multicolumn{1}{c}{$2\Delta n/n$} & \multicolumn{1}{c}{$\Delta V_{ES}$} &
     \multicolumn{1}{c}{$2\Delta n/n$} & \multicolumn{1}{c}{$\Delta V_{ES}$} \\
     (1/m$^3$) & \multicolumn{1}{c}{(nm)} &
     \multicolumn{1}{c}{($10^{-3}$)} & \multicolumn{1}{c}{(eV)} &
     \multicolumn{1}{c}{($10^{-3}$)} & \multicolumn{1}{c}{(eV)} \\
     \hline
     $10^{30}$ & 1.0  &26.0 & 1.31 & 4.17 & 0.069 \\
     $10^{30}$ & 2.0  &13.7 & 0.64 & 1.13 & 0.04  \\
     $10^{30}$ & 2.5  & 8.7 & 0.47 & 0.76 & 0.039 \\
     $10^{31}$ & 0.5  &20.5 & 2.75 & 1.27 & 0.087 \\
     $10^{31}$ & 1.0  & 9.5 & 1.27 & 0.71 & 0.073 \\
     $10^{31}$ & 1.25 & 7.6 & 1.66 & 0.37 & 0.112 \\
     $10^{32}$ & 0.25 & 1.9 &10.7  & 0.95 & 0.15  \\
     $10^{32}$ & 0.5  & 8.9 & 3.31 & 0.24 & 0.16  \\
     $10^{32}$ & 0.625& 7.2 & 4.23 & 0.27 & 0.25  \\
  \end{tabular*}
  \end{ruledtabular}
 \end{table}

\begin{figure}[ht]
   \begin{center}
   \includegraphics[width=\linewidth]{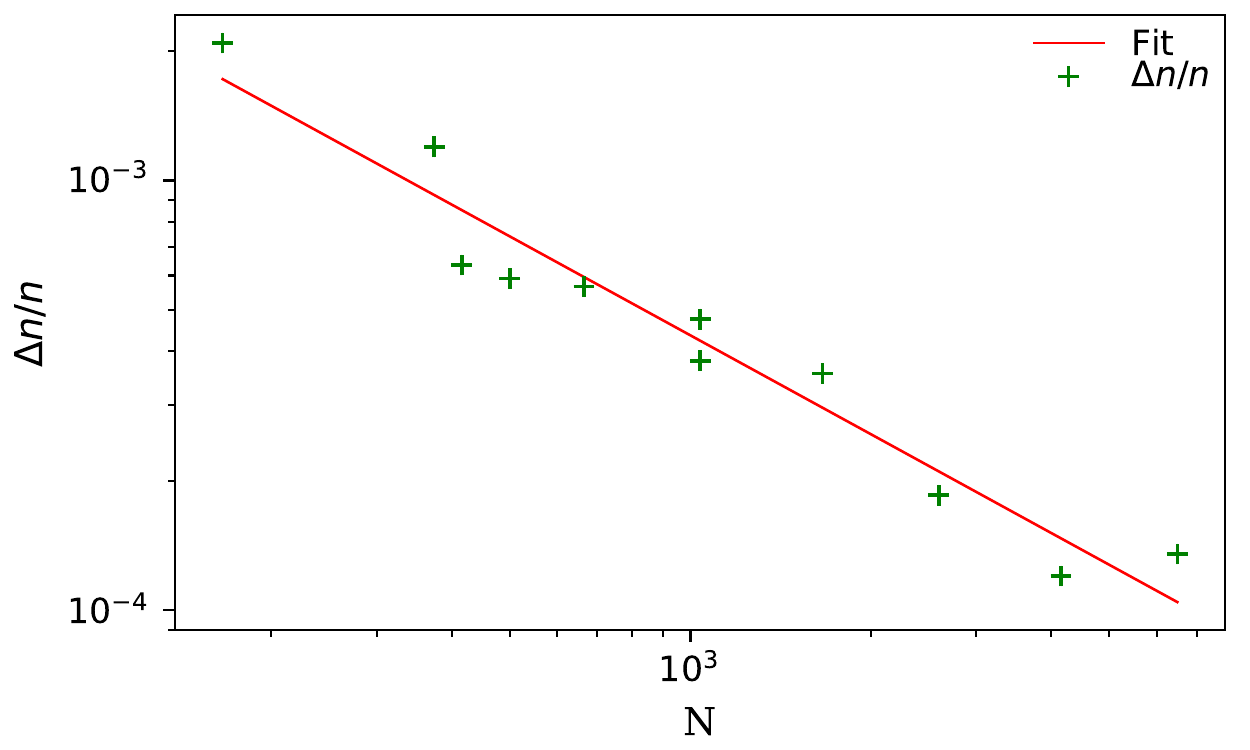}
   \caption{\label{fig:results_cylinder} Magnitudes of the electron density oscillations depending on the number of electrons $N$ in the logarithmic scale.}
   \end{center}
 \end{figure}

The results for amplitudes of potential oscillations and electron density oscillations are gathered in Table~\ref{tab:oscillation_amplitude_cylinder}.
As can be seen, the amplitude of the potential oscillations in an infinite cylinder is dramatically small compared to a sphere, in fact, the amplitude is about a factor of 20 smaller. The oscillations of the electron density in a cylinder are also less pronounced than in a sphere, and their dependence on electron number seems to be different.
This dependence is shown in Fig.~\ref{fig:results_cylinder}, which is similar to Fig.~\ref{fig:all_results}. The fitting line approximately corresponds to the law $\Delta n/n\propto N^{-0.76}\approx N^{-3/4}$.

Analyzing these results, we can conclude that presence of the spherical symmetry is essential for the effect, which is being considered in this paper.

 \begin{figure*}[p]
   \includegraphics[width=0.49\linewidth]{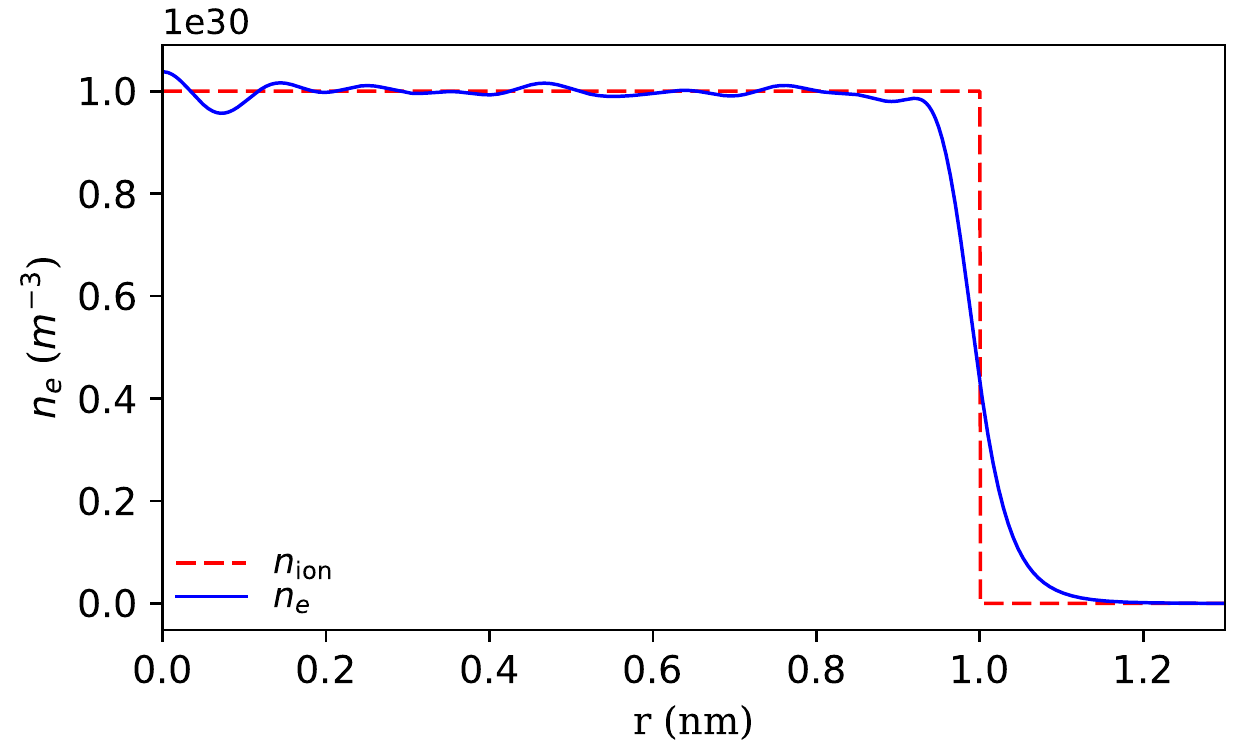}
   \includegraphics[width=0.49\linewidth]{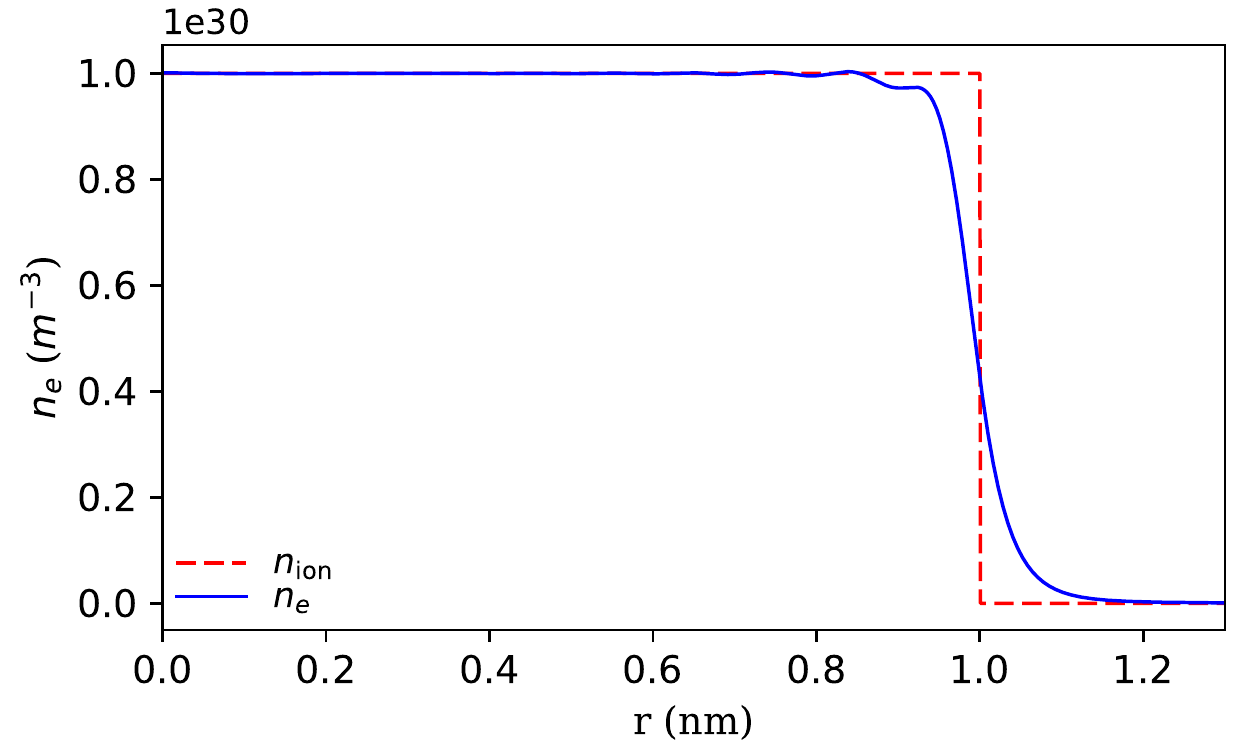}
   \caption{\label{fig:dens_with_temp_4188} The electron density in the bubble of radius
   1~nm with average density $10^{30} {\rm m}^{-3}$ (4188 electrons) when neglecting electron temperature (on the left) and with temperature 2~eV (on the right).}
 \end{figure*}

 \begin{figure*}[p]
   \includegraphics[width=0.48\linewidth]{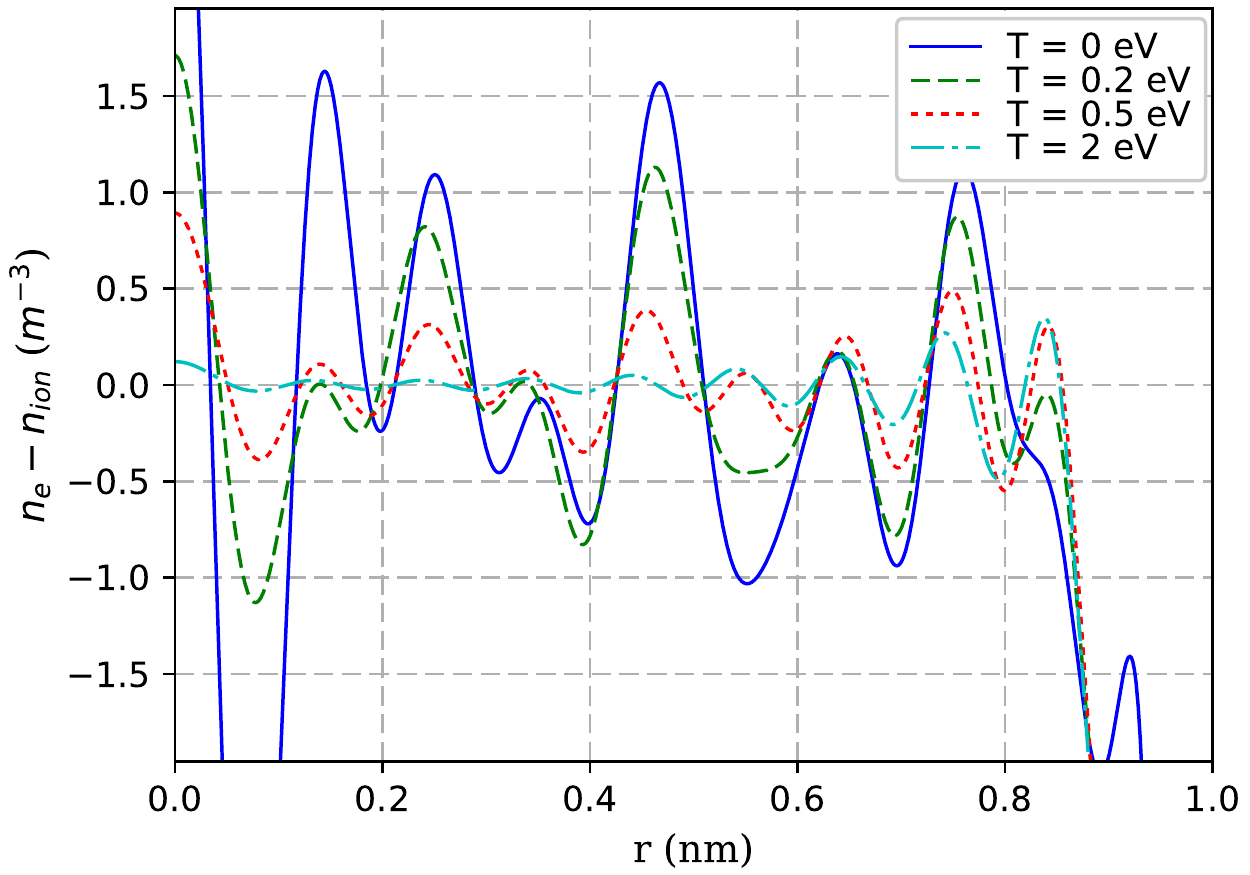}
   \includegraphics[width=0.48\linewidth]{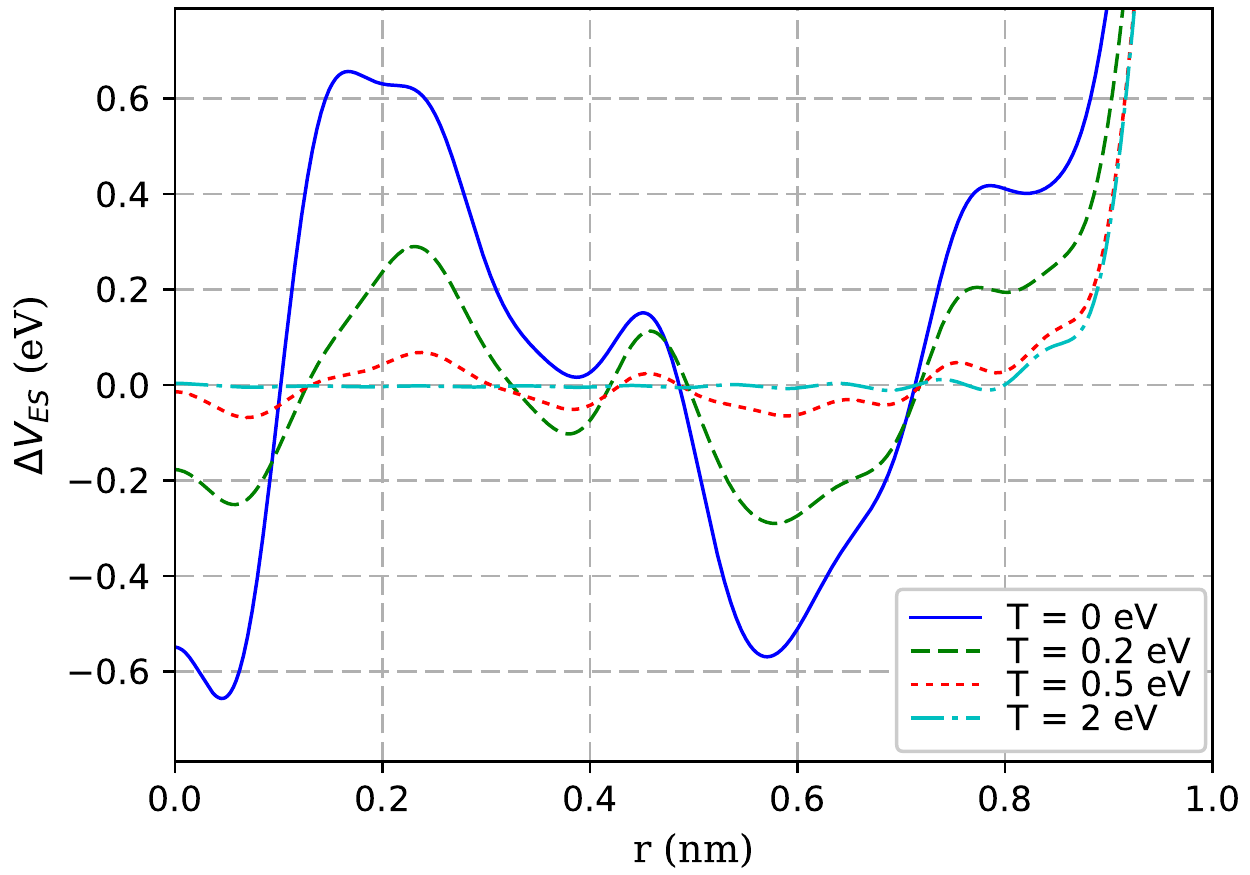}
  \caption{\label{fig:dens_osc_with_temp_4188} The electron density oscillations (on the left) and the electrostatic potential oscillations (on the right) in the bubble of radius 1~nm with
  the average density $10^{30} {\rm m}^{-3}$ (4188 electrons) at zero temperature of electrons, at electron temperature 0.2~eV, 0.5~eV, and 2~eV.}
 \end{figure*}

 \begin{figure*}[p]
   \includegraphics[width=0.48\linewidth]{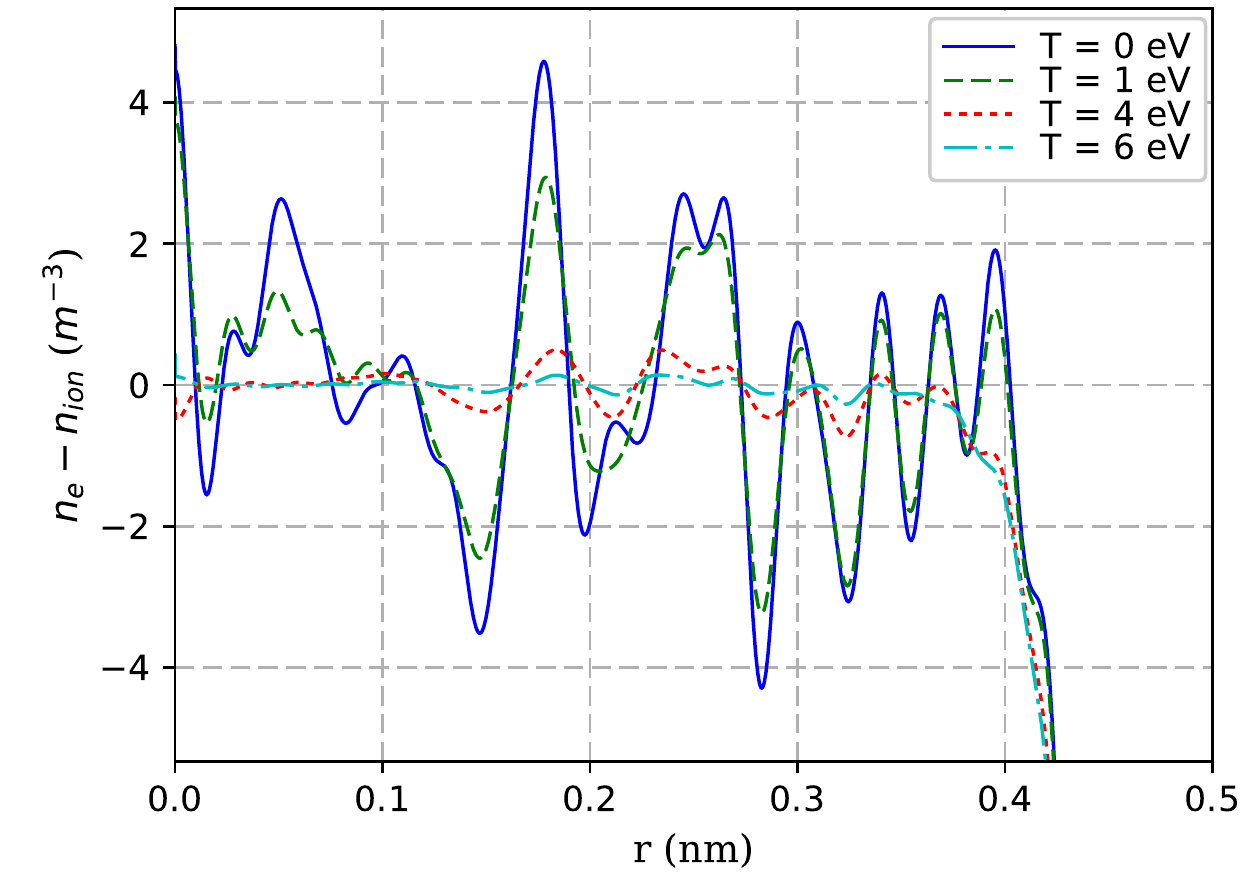}
   \includegraphics[width=0.48\linewidth]{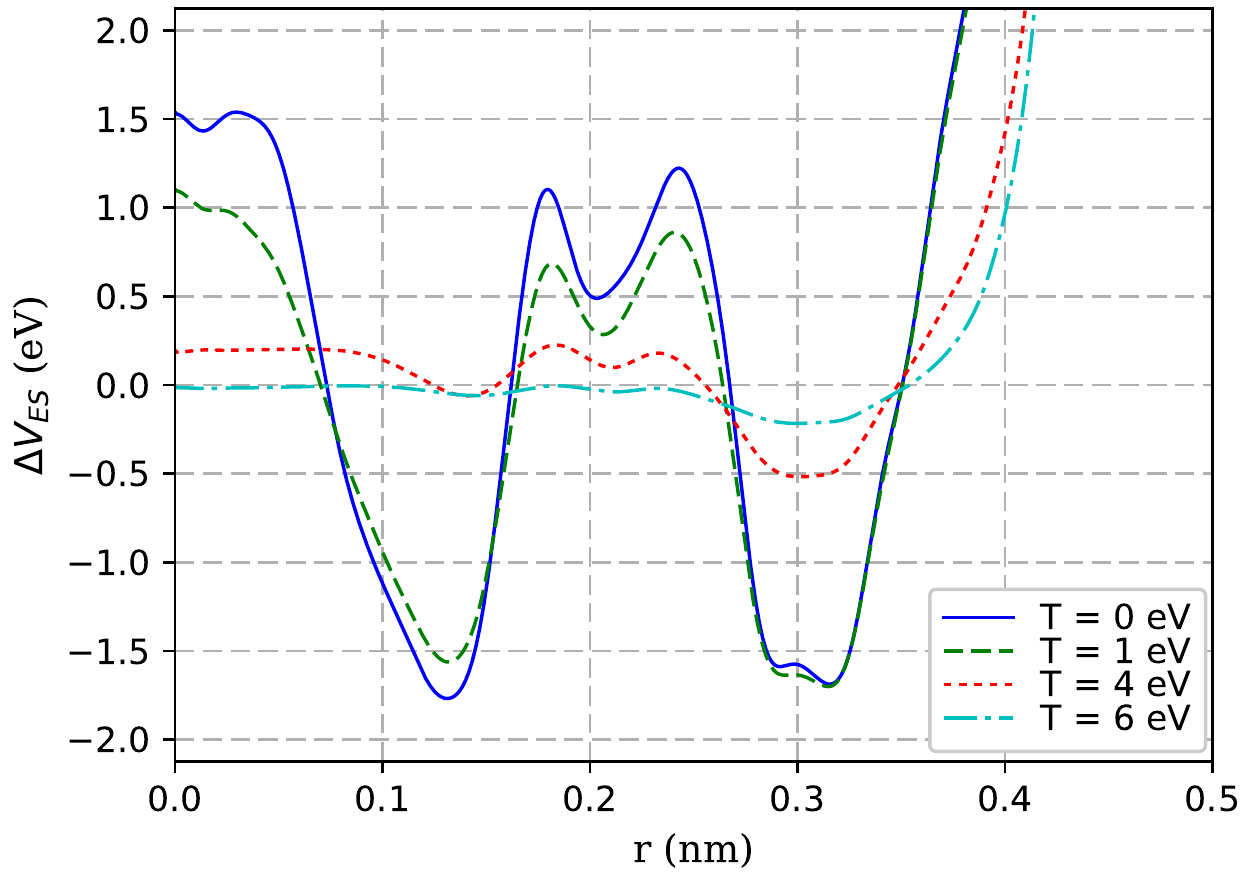}
  \caption{\label{fig:dens_osc_with_temp_52360} The electron density oscillations (on the left) and the electrostatic potential oscillations (on the right) in the bubble of radius 0.5 nm with average density $10^{32} {\rm m}^{-3}$ (52360 electrons) at zero temperature of electrons, at electron temperature 1 eV, 4 eV and 6 eV.}
 \end{figure*}

\subsection{Electron temperature}

Up to this point all results were obtained neglecting electron temperature. If we take into account electron temperature, then oscillations of the electron density and the electrostatic potential
diminish with an increase of temperature. To calculate the distribution of electrons with temperature $T>0$ we can use Eqs.~\eqref{Kohn-Sham} and~\eqref{density_sum}, but we should replace a step function $\Theta$ with the Fermi distribution~\cite{Engel}:
\begin{equation}
 \Theta_{nl}=\left[1+\exp\left(\dfrac{\varepsilon_{nl}-\mu}{k_{\rm B}T} \right) \right]^{-1},
\end{equation}
where $\mu$ is the chemical potential, determined from the condition $\sum \Theta_{nl}=N$.

Just to show the effect of electron temperature on the distribution of the electron density we made several calculations with electron temperature in a simple case of the uniform ionic jellium.
The electron density distribution in the bubble of radius 1~nm and with density $10^{30} {\rm m}^{-3}$ at zero electron temperature and at temperature 2~eV, is presented in Fig.~\ref{fig:dens_with_temp_4188}.
With the presence of electron temperature, the oscillations are much smaller and the density profile is almost flat. In Fig.~\ref{fig:dens_osc_with_temp_4188} the electron density and potential oscillations are shown in detail for the case mentioned above, with two more temperature values 0.2~eV and 0.5~eV.

Similar to Fig.~\ref{fig:dens_osc_with_temp_4188}, Fig.~\ref{fig:dens_osc_with_temp_52360} shows oscillations of the density and potential in the bubble of radius 0.5~nm with the average density $10^{32} {\rm m}^{-3}$ at several temperature values varying from zero to 6~eV.
In both cases the oscillations diminish with the temperature growth, keeping the shape of oscillations almost unchanged.

 All calculation results, presented in Table~\ref{tab:temperature_dependence}, were obtained for nonzero electron temperature.
 It is clear that the oscillation magnitudes of the density and potential decrease monotonically with the electron temperature growth.
 Moreover, the decrease is approximately exponential, and its rate is smaller at higher density, i.e., at bigger Fermi energy.

 \begin{table}[h]
  \caption{\label{tab:temperature_dependence} Dependence of the electron density oscillation amplitudes $2\Delta n/n$ and the potential oscillation amplitudes $2\Delta V_{\rm ES}$ on electron temperature $T$ for three values of average density and number of electrons.}
  \begin{ruledtabular}
   \begin{tabular}{D{.}{.}{1}D{.}{.}{1}D{.}{.}{3}D{.}{.}{1}D{.}{.}{2}D{.}{.}{3}D{.}{.}{1}D{.}{.}{3}D{.}{.}{3}}
    \multicolumn{3}{c}{\begin{tabular}{l}
                           $n=10^{30}m^{-3}$\\
                           $N=33510$\\
                          \end{tabular}
                          }
    &\multicolumn{3}{c}{\begin{tabular}{l}
                           $n=10^{31}m^{-3}$\\
                           $N=41888$\\
                          \end{tabular}
                          }
    &\multicolumn{3}{c}{\begin{tabular}{l}
                           $n=10^{32}m^{-3}$\\
                           $N=52360$\\
                          \end{tabular}
                          }\\
    \hline
    \multicolumn{1}{c}{T}&\multicolumn{1}{c}{$2\Delta n/n$}&\multicolumn{1}{c}{$\Delta V_{\rm ES}$}&
    \multicolumn{1}{c}{T}&\multicolumn{1}{c}{$2\Delta n/n$}&\multicolumn{1}{c}{$\Delta V_{\rm ES}$}&
    \multicolumn{1}{c}{T}&\multicolumn{1}{c}{$2\Delta n/n$}&\multicolumn{1}{c}{$\Delta V_{\rm ES}$}\\
    \multicolumn{1}{c}{(eV)} & \multicolumn{1}{c}{($10^{-3}$)} & \multicolumn{1}{c}{(eV)} &
    \multicolumn{1}{c}{(eV)} & \multicolumn{1}{c}{($10^{-3}$)} & \multicolumn{1}{c}{(eV)} &
    \multicolumn{1}{c}{(eV)} & \multicolumn{1}{c}{($10^{-3}$)} & \multicolumn{1}{c}{(eV)} \\
    \hline
     0  &13.7 & 0.64  &  0  & 9.5  & 1.27  &  0  & 8.9   & 3.31 \\
    0.1 &10.4 & 0.48  & 0.2 & 8.9  & 0.94  & 0.5 & 8.2   & 3.15 \\
    0.2 & 5.5 & 0.24  & 0.5 & 6.0  & 0.65  &  1  & 6.2   & 2.81 \\
    0.3 & 4.2 & 0.11  &  1  & 1.9  & 0.17  &  2  & 3.6   & 1.93 \\
    0.5 & 2.0 & 0.053 &  2  & 0.15 & 0.016 &  4  & 1.2   & 0.74 \\
    0.8 & 1.1 & 0.017 &     &      &       &  6  & 0.41  & 0.21 \\
        &     &       &     &      &       & 10  & 0.087 & 0.014 \\
   \end{tabular}
  \end{ruledtabular}
 \end{table}

\subsection{Curved boundary}

\begin{figure}[h]
  \centering
  \includegraphics[width=0.7\linewidth]{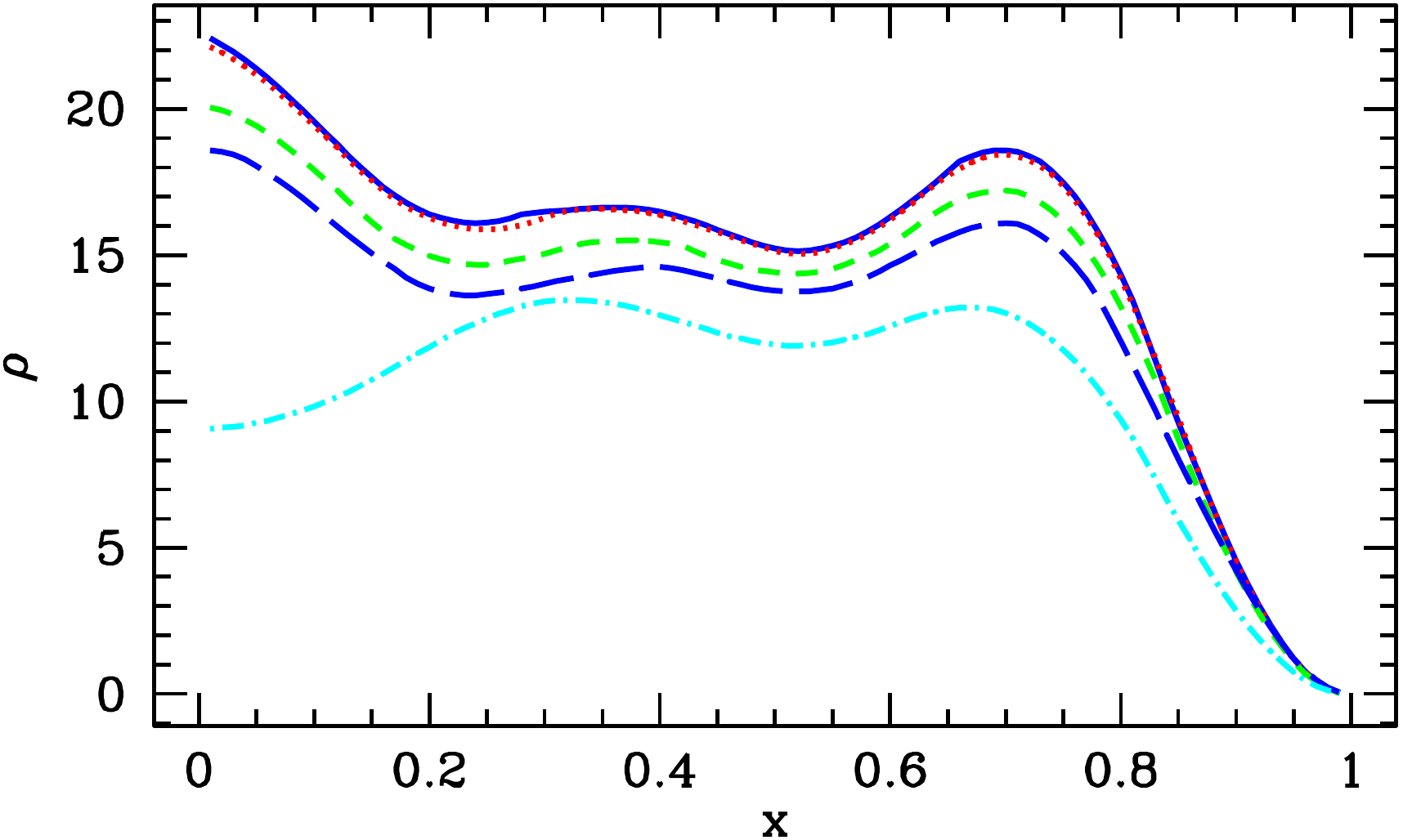}
  \caption{\label{N46} Oscillations of the electron density for the number of electrons $N_e=92$ in a spherical potential well (solid line); in a ellipsoidal potential well with ellipticity 1.01 (dotted line); with ellipticity 1.1 (short-dashed line); with ellipticity 1.2 (long-dashed line); with ellipticity 1.3 (dot-dashed line).}
\end{figure}

\begin{figure}[ht]
  \centering
  \includegraphics[width=0.7\linewidth]{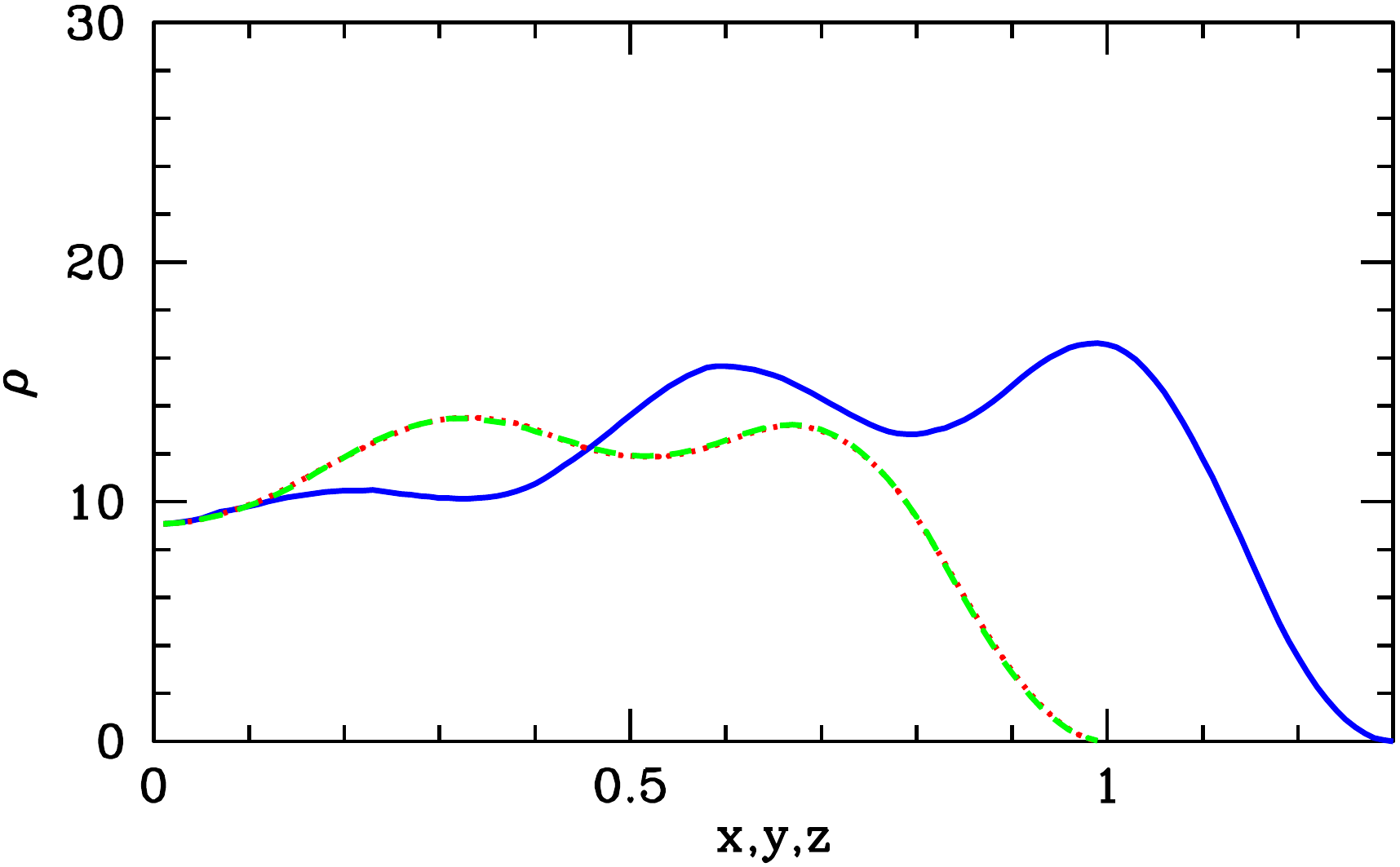}
  \includegraphics[width=0.7\linewidth]{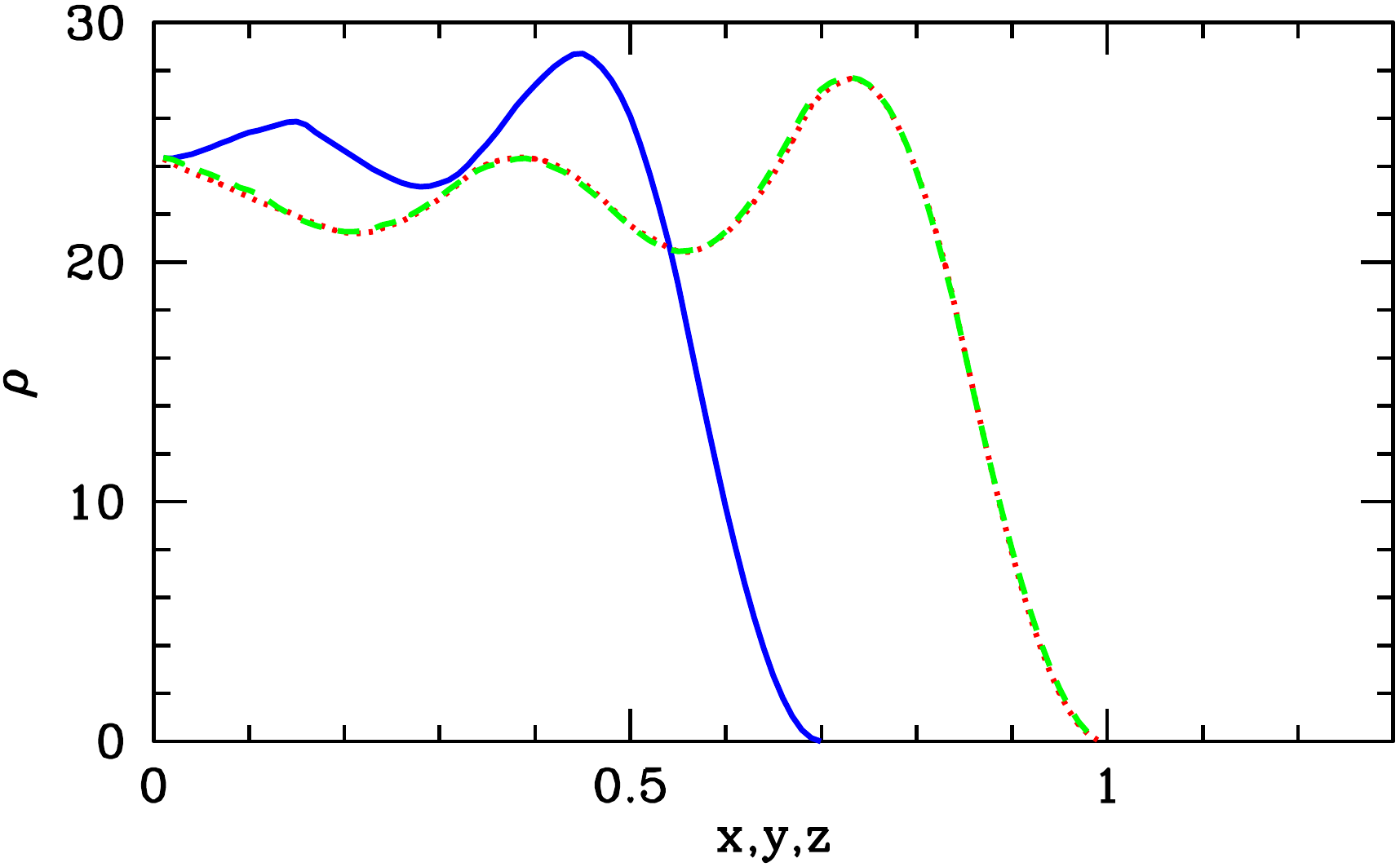}
  \caption{\label{N46xyz} Oscillations of the electron density for the number of electrons $N_e=92$ at ellipticity 1.3 (top) and 0.7 (bottom) along z axis. Solid lines show densities along z axis; dashed and dotted lines show densities along x and y axes, respectively, which are equal from symmetry.}
\end{figure}
All results described above are obtained in the strictly spherically symmetric approximation.
It is natural to check if the results persist for nonspherical configurations, and the effect of the oscillations does not disappear.
We have done several tests for the non-interacting electrons in the ellipsoidal potential well of
different ellipticity, in order to evaluate the influence of the curved boundary.
We employed package {\tt Mathematica 11.3} to perform those tests.
First, we checked how the finite-element solver for Laplace equation, embedded into
{\tt Mathematica 11.3}, reproduces the exact analytical solution for the spherical case.
We found that acceptable accuracy was obtained for $N_e<100$, for higher values of $N_e$
the errors become too large.

Here we present the distributions of the electron density in the ellipsoidal cavities.
We have tested several
values of the ellipsoid axes ratios (from 0.7 through 1.3), see Figs.~\ref{N46} and \ref{N46xyz}.

We may conclude that when the sphere is perturbed by a few percent, the perturbations of the
density are also of a few percent.
There is no qualitative difference between the ellipsoidal and spherical cases.
When perturbations grow up to 30\%, the distribution of electrons changes
dramatically (cf.~\cite{PhysRevE.96.042205} and references therein).

\section{Conclusion}

We discovered new nontrivial manifestation of the quantum shell effects in a mesoscopic system of degenerate electrons in a potential well.
It is shown that the electron distribution has a spatial scale of the order of system size.
This effect is confirmed by the theoretical analysis of the system of free degenerate electrons in an
infinite spherical well and by the numerical analysis of degenerate electrons in the compressed gas
bubble of submicron size.
The spatial distribution has several extreme points, and the amplitude of the deviation from the mean value depends on the number of electrons in the system
\begin{equation}
 \dfrac{\Delta n(r)}{n}\propto \left(\dfrac{8}{N}\right)^{1/2}.
\end{equation}
A consequence of the effect is the appearance of an electric field acting on the ion subsystem
\begin{equation}
 \Delta \varphi\propto \dfrac{eN^{1/2}}{\varepsilon_0R_0}.
\end{equation}
This leads to two nontrivial consequences.

Under conditions of hydrostatic equilibrium, the concentrations of electrons and ions in the compressed gas bubble equalize and coincide with the concentration of free electrons in the well.

The quantum shells effects are also manifested in the nontrivial dynamics of the compressible gas bubble. The hydrodynamics is fundamentally different from the process of adiabatic compression in traditional systems. The phenomenon of cumulation is observed in the system.

A number of factors limits and weakens the effect considered in the article. These factors have different origin.

The first is associated with the system symmetry. The numerical and theoretical analyses showed that the dependence of the relative deviation of the electron concentration in systems, possessing different symmetries, has a different nature
\begin{equation}
 \left(\frac{\Delta n}{n_0}\right)_{\rm sphere} \sim \frac{1}{\sqrt{N}},
\end{equation}
\begin{equation}
 \left(\frac{\Delta n}{n_0}\right)_{\rm cylinder} \sim \frac{1}{N^{3/4}},
\end{equation}
\begin{equation}
 \left(\frac{\Delta n}{n_0}\right)_{\rm flat} \sim \frac{1}{N}.
\end{equation}
For the macroscopic amounts of electrons ($N \sim 10^9$), these quantities can differ by several orders of magnitude.

The second factor is associated with the electron temperature. The calculation analysis showed, that at $T_e>0.1E_F$ the effect disappears.

The third factor is associated with the curved boundary. The numerical analysis showed that there is no qualitative difference between the spherical and ellipsoidal cases, when the sphere is perturbed by a few percent.

In conclusion, we note that the investigated inhomogeneity effect can be observed in the laboratory. The compressed gas bubble of submicron size is a realistic system. This system can be realized in the thermonuclear experiments.

\begin{acknowledgments}
 The authors are grateful to Y.E.~Lozovik for very helpful comments.
\end{acknowledgments}

\appendix
\section{Function $F_{\alpha}\left(\dfrac{r}{R_0},\dfrac{r}{R_0} \right )$}
\label{app_trajectories}
\begin{figure}[ht]
 \includegraphics[width=0.49\linewidth]{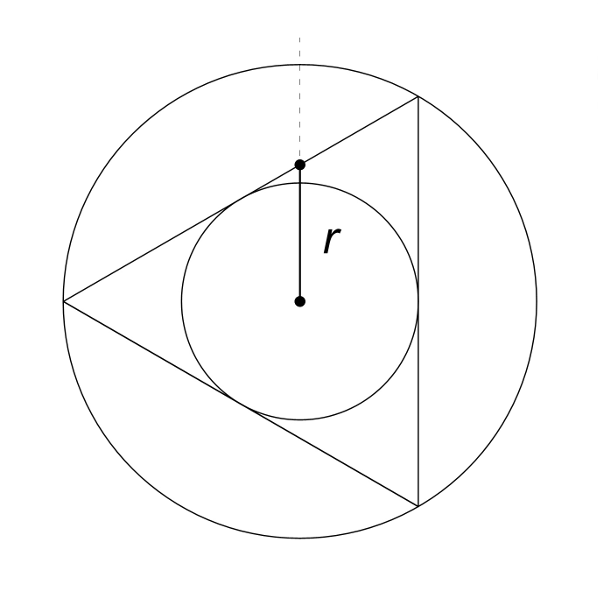}
 \includegraphics[width=0.49\linewidth]{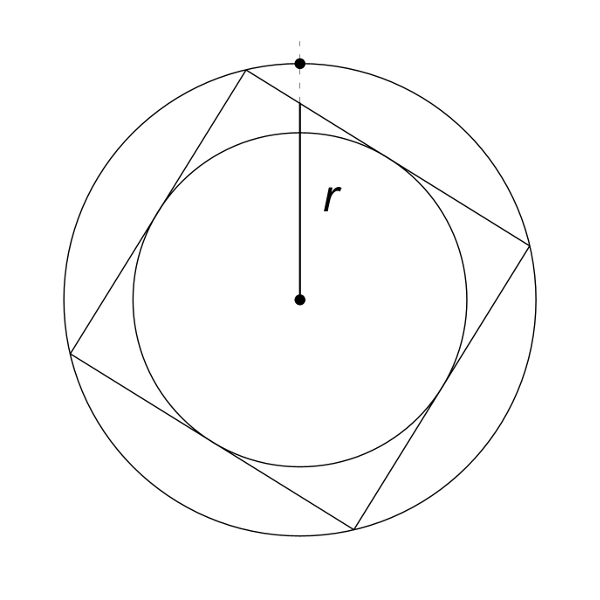}\\
 \hspace{0.03\linewidth}(a)\hspace{0.49\linewidth}(b)\\
 \caption{\label{fig:traj_rot} The trajectory of electron $\mbox{rot}(3,1)$~(a) and $\mbox{rot}(4,1)$~(b).}
\end{figure}

\begin{figure}[ht]
 \includegraphics[width=0.49\linewidth]{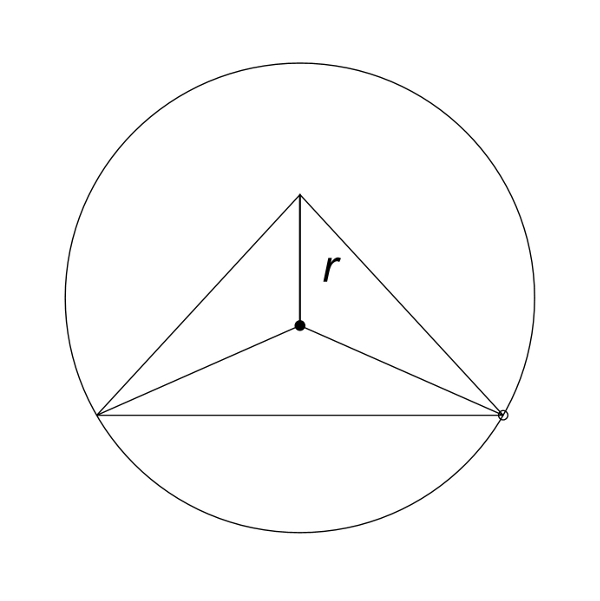}
 \includegraphics[width=0.49\linewidth]{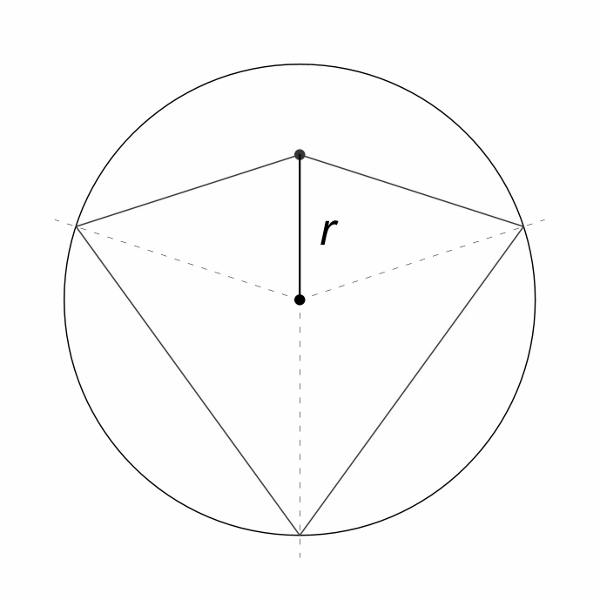}
 \hspace{0.03\linewidth}(a)\hspace{0.49\linewidth}(b)\\
 \caption{\label{fig:traj_def} The trajectory of electron $\mbox{def}(3,1)$~(a) and $\mbox{def}(4,1)$~(b).}
\end{figure}

We demonstrate the basic properties of the functions $F_{\alpha}\left(\dfrac{r}{R_0},\dfrac{r}{R_0} \right )$ on the example of the trajectories $\mbox{rot}(3,1)$, $\mbox{rot}(4,1)$ (Fig.~\ref{fig:traj_rot}) and $\mbox{def}(3,1)$, $\mbox{def}(4,1)$ (Fig.~\ref{fig:traj_def}). These trajectories give the main contribution to the sums~\eqref{delta_v_nonosc} and~\eqref{delta_v_osc}, respectively, since trajectories with small values of $m$ give the main contribution to the value of $\Delta n(r)$. This is due to the fact that each term of the sum~\eqref{delta_n} is proportional $1/L_{\alpha}\left(\dfrac{r}{R_0},\dfrac{r}{R_0}\right)$ and decreases with increasing $m$, since $L_{\alpha}\sim m$.

For these trajectories it is possible to obtain closed analytical expressions for $F_{\alpha}$ (there we denote $x=r/R_0$). For trajectories of $\mbox{rot}$ type we have
\begin{gather}
 F_{\mbox{rot}(3,1)}(x,x)=\begin{cases}
                    -\dfrac{1}{2x}\sqrt{\dfrac{1}{\sqrt{3}(4x^2-1)}},\quad &\mbox{at }x\geq1/2,\\
                    0,\quad &\mbox{at }x<1/2,
                   \end{cases}\\
 L_{\mbox{rot}(3,1)}(x,x)=3\sqrt{3},
\end{gather}
and
\begin{gather}
 F_{\mbox{rot}(4,1)}(x,x)=\begin{cases}
                    \dfrac{1}{4\sqrt{2}(2x^2-1)},\quad &\mbox{at }x\geq\sqrt{2}/2,\\
                    0,\quad &\mbox{at }x<\sqrt{2}/2,
                   \end{cases}\\
 L_{\mbox{rot}(4,1)}(x,x)=4\sqrt{2}.
\end{gather}
For trajectories of $\mbox{def}$ type we have

\begin{widetext}
\begin{gather}
 F_{\mbox{def}(3,1)}(x,x)=\dfrac{-1+\sqrt{8x^2+1}}{4x^2} \sqrt{\dfrac{8x^3(4x^2+5+3\sqrt{8x^2+1})} {\sqrt{8x^2+1}\left(4x^2-1+\sqrt{8x^2+1}\right)^{5/2} \sqrt{4x^2+1+\sqrt{8x^2+1}}}}, \quad \mbox{at }x\geq0,\\
 L_{\mbox{def}(3,1)}(x,x)=3\sqrt{3},
\end{gather}
\begin{gather}
 F_{\mbox{def}(4,1)}(x,x)=\begin{cases}
                    \sqrt{\dfrac{\sqrt{\dfrac{x}{(1-2x)^2(1+x)}}}{(3x-1)}} \sin\left[3\arccos\dfrac{1}{2}\sqrt{1+\dfrac{1}{x}}\right], \quad &\mbox{at }x\geq1/3,\\
                    0,\quad &\mbox{at }x<1/3,
                   \end{cases}\\
 L_{\mbox{def}(4,1)}(x,x)=3\sqrt{3},
\end{gather}
\end{widetext}

The number of important properties of the function $F_{\alpha}$ are derived from the obtained expressions:
\begin{enumerate}
 \item The function $F_{\alpha}$ is nonzero for $r>s_{\alpha}$
 \begin{enumerate}
  \item $s_{\mbox{rot}(n,1)} = R_0\sin\dfrac{(n-2)\pi}{2n}$.
  \item $s_{\mbox{def}(n,1)} \cong s_{\mbox{rot}(n-1,1)}$.
 \end{enumerate}
 \item The function $F_{\alpha}$ has singular points $d_{\alpha}$
 \begin{enumerate}
  \item $d_{\mbox{rot}(n,1)} = s_{\mbox{rot}(n,1)}$.
  \item $d_{\mbox{def}(n,1)} = s_{\mbox{rot}(n-1,1)}$.
 \end{enumerate}
\end{enumerate}
So for $F_{\alpha}$ we have following expression
\begin{equation}
 F_{\alpha}\left(\frac{r}{R_0},\frac{r}{R_0} \right)=\Theta(r-s_{\alpha}) G_{\alpha}\left(\frac{r}{R_0},\frac{r}{d_{\alpha}} \right).
\end{equation}

We note, that after summing up everything, the singularities are compensated.

\section{Stabilized jellium model}
\label{app_jellium}

In the case of spherically symmetric charge distribution, the Poisson equation $\Delta \varphi=4\pi n(r)$ can be simply integrated and for the electron electrostatic potential $V_H$, we would get
\begin{equation}
  \label{V_H_formula}
  V_{H}(r)=\frac{1}{r}\left(\int_0^r4\pi s^2n_e(s){\rm d}s\right) + \int_r^{+\infty}4\pi sn_e(s){\rm d}s .
\end{equation}
The potential equals zero at infinity. If ionic jellium density is uniform inside the radius $R$, then potential of ions $V_{\rm ion}$ has a simple form
\begin{equation}
  \label{homogeneous_core}
  V_{\rm ion}(r)=Z\begin{cases}
                 -\dfrac{1}{r},&r>R\\
                 \dfrac{1}{2R}\left(\dfrac{r^2}{R^2}-3\right),&r\le R
                \end{cases} .
\end{equation}
In the other case it can be calculated similar to $V_H$ using~\eqref{V_H_formula}, in which $n_e$ is replaced by $-n_{\rm ion}$.

In the stabilized jellium model there is a correction for the effective Kohn-Sham potential $\langle\delta v\rangle_{WS}$, that is the difference between the potential of the uniformly charged ball with ion charge $V(r)$ and the model pseudopotential averaged over the volume per ion $Z/n_{\rm ion}$~\cite{PerdewSJ, PerdewSJ2, Kiejna}
\begin{equation}
 \langle\delta v\rangle_{WS}=\frac{3}{4\pi r_0^3}\int_0^{r_0}{\rm d}r4\pi r^2[\omega(r)+V(r)],
\end{equation}
\begin{equation}
 r_0=Z^{1/3}r_s=Z^{1/3}\left(\frac{3}{4\pi n_{\rm ion}}\right)^{1/3},
\end{equation}
where $Z$ is the ion charge, $n_{\rm ion}$ is the ionic jellium density, $V(r)$ is the potential of the homogeneously charged ball, $\omega(r)$ is the model pseudopotential of interaction between an electron and an ion with charge $Z$:
\begin{equation}
 \label{Ashcroft}
 \omega(r)=\begin{cases}
       -Z/r&, r>r_c\\
       0&, r<r_c.
      \end{cases}
\end{equation}
Some comparison of the stabilized jellium model with three-dimensional calculations is presented in~\cite{AFM1}.

\section{Solution for an infinite cylinder}
\label{app_cylinder}

When the external potential has axial symmetry, and it is constant along symmetry axis (we name it $z$), single electron wavefunctions can be decomposed as
\begin{equation}
  \label{cylindrical_decomp}
 \phi(r,\varphi,z)=\frac{P(r)}{\sqrt{r}}e^{\pm im\varphi}e^{ip_z z}.
\end{equation}
Substituting single electron wavefunctions for the Kohn-Sham equation
\begin{equation}
 \hat{H}\phi=\left(-\frac{1}{2}\Delta+V_{\rm KS} \right)\phi=E\phi,
\end{equation}
we get
\begin{equation}
 \left(-\frac{1}{2}\left(\frac{{\rm d}^2}{{\rm d}r^2}-\frac{m^2-1/4}{r^2}-p_z^2\right)+V_{\rm KS} \right)P(r)=EP(r).
\end{equation}
If we denote $E_{nm}=E-p_z^2/2$, multiply the equation by $-2$ and move all terms to the left side, we finally get equation
\begin{equation}
 \left(\frac{{\rm d}^2}{{\rm d}r^2}-\frac{m^2-1/4}{r^2}+2[E_{nm}-V_{\rm KS}] \right)P_{nm}(r)=0,
\end{equation}
which is similar to Eq.~\eqref{Kohn-Sham}, if we make a replacement $l\rightarrow m-1/2$.

At zero temperature, $E\leq E_{\rm F}$ for all occupied states, so momentum component $p_z$ of the electron with the radial wavefunction $P_{nm}(r)$ satisfies a relation
\begin{equation}
 \frac{p_z^2}{2}=E-E_{nm}\leq E_{\rm F}-E_{nm},
\end{equation}
i.e., momentum $p_z$ lies in the range
\begin{equation}
 -\sqrt{2(E_{\rm F}-E_{nm})} \leq p_z \leq +\sqrt{2(E_{\rm F}-E_{nm})}.
\end{equation}
Let us denote the number of electrons per length unit of a cylinder as $\tau={\rm d}N/{\rm d}z$, and
the respective number in states with quantum numbers $n$ and $m$ as $\tau_{nm}$. The number of electrons equals the phase volume, divided by the volume of one state ($\hbar=1$)
\begin{equation}
 {\rm d} N_{nm}=g_eg_m\frac{{\rm d} p_z{\rm d} z}{2\pi \hbar}=g_eg_m\frac{2\sqrt{2(E_{\rm F}-E_{nm})}{\rm d} z}{2\pi},
\end{equation}
where $g_e=2$ is spin degeneration factor, $g_m$ is a factor, arising from the sign in $e^{\pm im\varphi}$, $g_m=1$ at $m=0$ and $g_m=2$ at $m\neq0$ i.e.,
\begin{equation}
 \tau_{nm}=\frac{{\rm d} N_{nm}}{{\rm d} z}=g_m\frac{2\sqrt{2(E_{\rm F}-E_{nm})}}{\pi}.
\end{equation}
For the full number of electrons per unit length, the following expression is correct
\begin{equation}
 \tau=\sum_{E_{nm}\leq E_{\rm F}}\tau_{nm}=\sum_{E_{nm}\leq E_{\rm F}}g_m\frac{2\sqrt{2(E_{\rm F}-E_{nm})}}{\pi},
\end{equation}
so expression for the electron density has a form
\begin{equation}
 n_e(r)=\sum_{E_{nm}\leq E_{\rm F}}g_m\frac{P_{nm}^2}{2\pi r}\frac{2\sqrt{2(E_{\rm F}-E_{nm})}}{\pi},
\end{equation}
on condition that $P_{nm}(r)$ is normalised to unit.

The Fermi energy value $E_{\rm F}$ can be determined numerically, for example, using binary search between maximum energy value $E_{nm}$, that gives $\tau(E_{nm})<\tau$ and the next closest value.

The electrostatic potential of a cylindrical surface with radius $R$ is
\begin{equation}
 V_{\rm surf}(r;R,\tau)=\begin{cases}
           0,&r\leq R\\
           -2\tau \ln\dfrac{r}{R},&r>R.
          \end{cases}
\end{equation}
As potential indefinitely grows at infinity, it is convenient to take the cylinder center $r=0$ as a zero of the potential.
The potentials $V_{\rm H}$ and $V_{\rm ion}$, similar to~\eqref{V_H_formula},~\eqref{homogeneous_core} (also multiplied by -1), are equal to
\begin{equation}
\begin{split}
 V_{\rm H}(r)=\int_0^r V_{\rm surf}(r;s,n_e(s))2\pi s{\rm d}s=\\
 =-\int_0^r 4\pi sn_e(s)\ln\frac{r}{s}{\rm d}s,
 \end{split}
\end{equation}
\begin{equation}
 V_{\rm ion}(r)=\begin{cases}
              \pi n_i r^2,&r\leq R\\
              \pi n_i R^2\left(1+2\ln\dfrac{r}{R} \right),&r>R
             \end{cases}
\end{equation}
for the uniform ionic density inside the radius $R$.

\bibliographystyle{apsrev}
\bibliography{References}

\end{document}